\documentclass[twocolumn]{aastex62}

\usepackage{rotating}
\usepackage{subfigure}
\usepackage{longtable}
\usepackage{url}
\usepackage{morefloats}
\usepackage{amsmath} 
\usepackage{tikz}
\usetikzlibrary{shapes,arrows}
\usepackage{amsmath}
\usepackage{amssymb}

\newcommand{\Kepler}{{\it Kepler~}}

\newcommand{\teff}{$T_{\rm eff}$}

\newcommand\aastex{AAS\TeX}

\received{September 21, 2017}
\revised{February 10, 2018}
\accepted{February 19, 2018}
\submitjournal{The Astronomical Journal}

\shorttitle{\aastex\ TESS Cool Dwarfs}
\shortauthors{Muirhead et al.}

\begin{document}

\title{A Catalog of Cool Dwarf Targets for the Transiting Exoplanet Survey Satellite}

\correspondingauthor{Philip S. Muirhead}
\email{philipm@bu.edu}

\author[0000-0002-0638-8822]{Philip S. Muirhead}
\affiliation{Department of Astronomy, Institute for Astrophysical Research, Boston University, 725 Commonwealth Avenue, Boston, MA 02215, USA}

\author[0000-0001-8189-0233]{Courtney Dressing}
\affiliation{Department of Astronomy, The University of California, Berkeley, CA  94720, USA}

\author[0000-0003-3654-1602]{Andrew W. Mann}
\altaffiliation{Hubble Fellow}
\affiliation{Department of Astronomy, The University of Texas at Austin, Austin, TX 78712, USA}

\author[0000-0002-0149-1302]{B\'{a}rbara Rojas-Ayala}
\affiliation{Departamento de Ciencias Fisicas, Universidad Andres Bello, Fernandez Concha 700, Las Condes, Santiago, Chile}


\author{Sebastien Lepine}
\affiliation{Department of Physics and Astronomy, Georgia State University, Atlanta, GA 30303 USA}

\author[0000-0001-8120-7457]{Martin Paegert}
\affiliation{Harvard-Smithsonian Center for Astrophysics, 60 Garden St, Cambridge, MA 02138 USA}

\author{Nathan De Lee}
\affiliation{Department of Physics, Geology and Engineering Tech, Northern Kentucky University, Highland Heights, KY 41099, USA}
\affiliation{Department of Physics \& Astronomy, Vanderbilt University, 6301 Stevenson Center Ln., Nashville, TN 37235, USA}

\author[0000-0002-0582-1751]{Ryan Oelkers}
\affiliation{Department of Physics \& Astronomy, Vanderbilt University, 6301 Stevenson Center Ln., Nashville, TN 37235, USA}

\begin{abstract}

We present a catalog of cool dwarf targets ($V-J>2.7$, $T_{\rm eff} \lesssim 4000 K$) and their stellar properties for the upcoming Transiting Exoplanet Survey Satellite (TESS), for the purpose of determining which cool dwarfs should be observed using two-minute observations.  TESS has the opportunity to search tens of thousands of nearby, cool, late K and M-type dwarfs for transiting exoplanets, an order of magnitude more than current or previous transiting exoplanet surveys, such as {\it Kepler}, K2 and ground-based programs.  This necessitates a new approach to choosing cool dwarf targets.  Cool dwarfs were chosen by collating parallax and proper motion catalogs from the literature and subjecting them to a variety of selection criteria.  We calculate stellar parameters and TESS magnitudes using the best possible relations from the literature while maintaining uniformity of methods for the sake of reproducibility.  We estimate the expected planet yield from TESS observations using statistical results from the {\it Kepler} Mission, and use these results to choose the best targets for two-minute observations, optimizing for small planets for which masses can conceivably be measured using follow up Doppler spectroscopy by current and future Doppler spectrometers.  The catalog is incorporated into the TESS Input Catalog and TESS Candidate Target List until a more complete and accurate cool dwarf catalog identified by ESA's Gaia Mission can be incorporated.

\end{abstract}


\section{Introduction}

Cool dwarf stars, specifically late K dwarf and M dwarf stars, are exciting targets for exoplanet surveys.  Compared to sun-like and earlier-type dwarfs, the smaller masses and radii of cool dwarf stars enable the detection and characterizing of smaller and less-massive exoplanets via the transit and radial velocity techniques \citep[e.g.][]{Nutzman2008, Muirhead2011}.  For these reasons, some of the smallest exoplanets found to date orbit cool dwarf stars, including sub-Earth-sized exoplanets, such as {\it Kepler}-1308 b (0.51 $R_{\Earth}$), {\it Kepler}-138 b \citep[0.52 $R_{\Earth}$, both from][]{Morton2016}, K2-89 b \citep[0.62 $R_{\Earth}$,][]{Crossfield2016} and {\it Kepler}-42 c \citep[0.73 $R_{\Earth}$,][]{Muirhead2012a,Mann2017}.  Planets orbiting within cool dwarf stars' habitable zones are more easily discovered than those orbiting within the habitable zones of sun-like stars.  Moreover, potentially-habitable exoplanets orbiting cool dwarfs are more easily characterized via transit transmission spectroscopy, thanks to the increased number of transits in a given amount of time and the relatively deep transit signals from terrestrial-size planets \citep[][]{Kaltenegger2009,Belu2011}.  In fact, a recent study by \citet[][]{Kane2016} found that of all the planet candidates discovered by NASA's {\it Kepler} Mission that are less than 2.0 $R_\Earth$ and reside within an optimistically-sized habitable zone, 40\% orbit stars with effective temperatures less than 4000 K.  This, despite the fact that cool dwarfs make up less than 5\% of the initial {\it Kepler} target sample \citep[][]{Batalha2010}.  Investigations of planets orbiting M dwarfs show that the majority of M dwarfs host greater than 2 planets with periods of less than 200 days \citep[e.g.][]{Dressing2013,Gaidos2014}, that one in five mid-M dwarfs host compact multiple systems \citep[][]{Muirhead2015}, and that one in seven M dwarfs host an Earth-sized planet orbiting within the habitable zone \citep[][]{Dressing2015}.

Because they both outnumber and are intrinsically fainter than sun-like stars, bright (J$<$12) cool dwarfs tend to be more evenly distributed across the sky, rather than concentrated toward the galactic plane.  For this reason, NASA's {\it Kepler} and K2 Missions can only observe a few thousand cool dwarf stars continuously for transiting exoplanets.  Ground-based transit programs, such as MEarth \citep[e.g.][]{Berta2013} and TRAPPIST \citep[e.g.][]{Gillon2012} monitor the brightest and nearest cool dwarfs by individually targeting them one at a time.  However, these programs suffer from noise associated with ground-based precision photometry and diurnal and weather-induced time-coverage challenges, similarly limiting their target lists to hundreds or thousands.  In contrast, NASA's upcoming Transiting Exoplanet Survey Satellite (TESS) is uniquely suited to search an order of magnitude more cool dwarfs for transiting exoplanets by utilizing wide-angle imaging cameras in a space environment.  

\citet[][]{Ricker2014} provided a detailed description of the observing mode of TESS.  Similar to {\it Kepler}, TESS will have a ``two-minute'' observing mode, wherein 2-minute exposures are acquired for a limited number of apertures in the TESS fields.  So-called ``full-frame'' observations, consisting of 30-minute exposures, are acquired for the entirety of each TESS field.  The apertures chosen for two-minute observations require careful consideration.  \citet[][hereafter S15]{Sullivan2015} first calculated the number of stars of various spectral types that TESS could observe with two-minute observations in order to maximize the number of planet discoveries, using statistics from prior exoplanet surveys.  They showed that optimally, TESS would observe roughly 50,000 stars with effective temperatures less than 4,000 K and TESS magnitudes brighter than 16, and should detect roughly 500 transiting planets orbiting those stars.

However, S15 did not use literature star catalogs in their simulation.  Instead, they used galactic models to simulate observable stars.  Determining the actual 50,000 cool dwarf stars that TESS should observe is itself a challenge.  By far the most reliable method for identifying individual cool dwarfs is with archival trigonometric parallax observations via mass-luminosity relations.  Trigonometric parallax measurements provide absolute magnitudes for stars, and absolute infrared magnitude has been shown to determine star mass with no perceivable effect from stellar metallicity \citep[e.g.][]{Henry1993, Henry1999, Delfosse2000, Boyajian2012, Mann2013c, Benedict2016}.  Unfortunately, archival trigonometric parallax measurements for stars, such as those measured by the {\it Hipparcos} Mission \citep[][]{vanleeuwen2007}, do not include significant numbers of cool dwarfs due to their intrinsic faintness.  Soon, ESA's Gaia Mission will measure trigonometric parallaxes for hundreds of thousands of cool dwarfs, and those measurements will be enormously useful for deciding the TESS targets appropriate for 2-minute cadence.  In the meantime, however, TESS cool dwarf targets must be chosen by other means.

In the absence of trigonometric parallaxes, cool dwarfs must be selected using archival spectroscopic, color and/or proper motion measurements.  In this paper, we describe a catalog of cool dwarfs for two-minute TESS observations using  archival parallaxes where available, or proper motion observations from the SUPERBLINK program \citep[][]{Lepine2011}, in combination with a variety of photometric catalogs.  In Section \ref{sec:methods}, we discuss our methods for identifying cool dwarf stars, and in Section \ref{sec:yield}, we compare the expected planet yields for TESS observations of real stars to those of S15.  In Section \ref{sec:discussion}, we discuss the importance of this catalog of cool dwarfs for TESS discoveries and follow up observations.

\section{Methods}\label{sec:methods}

\subsection{Catalogs}

Despite the emergence of high volume, large-scale sky surveys, all-sky star catalogs that include tens of thousands of bright cool dwarfs are surprisingly rare.  Spectroscopic surveys either do not contain the necessary number of cool dwarfs to meet the simulations from S15, or the stars in the catalogs are too faint to be good TESS targets ($I_C>15$).  \citet{West2008} presented over 70,000 M dwarfs spectroscopically verified from the Sloan Digital Sky Survey (SDSS); however the fields observed by SDSS are limited to specific regions of the sky with visible-band magnitudes greater than 15.  The same is true for cool dwarfs with spectra measured by LAMOST \citep[][]{Yi2014}.  S15 calculated that cool dwarfs fainter than 15th magnitude in $I$-band are not ideal for TESS observations due to the significant role of photon noise.

On the other hand, {\it targeted} spectroscopic surveys of nearby cool dwarfs, such as the Palomar/Michigan State University (PMSU) Survey \citep[][]{Reid1995, Hawley1997, Gizis2002, Reid2002} are nearly all sky, but only include hundreds of cool dwarfs.  More recently,  cool dwarf spectroscopic surveys by \citet[][]{Rojas2012}, \citet[][]{Deshpande2013}, \citet{Newton2014},  \citet{Terrien2015} and \citet[][]{Zhong2015} have increased the number of spectroscopically characterized bright cool dwarfs.  However, together, these spectroscopic surveys have characterized only about 2000 cool dwarfs, not the tens of thousands needed to optimally assign two-minute apertures for TESS observations.

In order to acquire tens of thousands of cool dwarfs with TESS magnitudes brighter than $T$ of 16, we must turn to photometric surveys and select objects based on broadband colors alone.  The Two-Micron All-Sky Survey \citep[2MASS][]{Cutri2003, Skrutskie2006} provides $J$, $H$ and $K_s$-band magnitudes for nearly all stars that TESS can observe for transiting planets.  $J$, $H$ and $K_s$-band colors provide some information on the properties of stars, and can be effectively used for isolating late M dwarfs from earlier-type stars and, in some cases, evolved stars. However, on their own $J$, $H$ and $K_s$-band colors are of limited use for identifying late K or early M dwarf stars, due to their significant overlap with evolved stars \citep[see for example][their Figure 5]{Bessell1988} and with earlier-type stars suffering moderate amounts of interstellar reddening.  Similarly, the Wide-field Infrared Survey Satellite (WISE) provides mid-infrared magnitudes for the entire sky; although for typical M dwarf temperatures, there is little difference between 2MASS magnitudes and WISE magnitudes, both bands being on the Rayleigh-Jeans side of the Planck Law corresponding to these effective temperatures.

The best method to identify M dwarfs in photometric surveys is to include proper motion data. Stars with moderately large proper motions ($>20$ mas yr$^{-1}$) tend to be relatively nearby, and this presents two advantages. On the one hand, red giants and red dwarfs have large ($\gtrsim5$ mag) absolute magnitude differences, and giants can thus be easily identified in samples of high proper motion stars, as they are systematically brighter. On the other hand, stars with large proper motions are also relatively nearby ($d<$1 kpc) and thus unlikely to suffer significant amounts of reddening. As a result, samples of stars with large proper motions are generally dominated by nearby K and M dwarfs, and their effective temperatures can be estimated with some certainty based on broadband, optical-to-infrared colors.

The SUPERBLINK proper motion catalog (Lepine, {\it in prep}), is an all-sky catalog of stars with large proper motions ($\mu>40$ mas yr$^{-1}$) which includes optical and infrared magnitudes for all its entries. These include optical G-band magnitudes from the first GAIA release, whenever available, and estimated optical V-band magnitudes for all the stars. The SUPERBLINK survey is based out of a project to identify all high proper motion stars from archival plates from the Digitized Sky Surveys (DSS). SUPERBLINK uses, for example, images from the National Geographic Palomar Observatory Sky Survey \citep[POSS I][]{Minkowski1963} and the Second Palomar Observatory Sky Survey \citep[POSS II, conducted roughly 40 years later,][]{Reid1991}.  Both surveys were originally performed using photograph plates, and both have since been digitized \citep[][]{Djorgovski2002,Gal2004,Odewahn2004}, enabling computation methods for identifying high proper motion objects.

Instead of directly measuring stellar positions, SUPERBLINK use an image-differencing algorithm, which identifies moving objects from their patterns of residuals after image subtraction. As a result, the method works successfully in crowded fields of low Galactic latitudes \citep[see e.g.][]{Lepine2002}. In addition, the SUPERBLINK survey includes stringent quality control tests, and all but the most obvious detections were individually examined by eye, on the computer screen, using a blink-comparator widget. Additional processing includes cross-correlation with several photometric catalogs, including GALEX, SDSS, USNO-B1.0, GAIA DR1,  2MASS, and WISE. The proper identification of the counterparts of these high proper motion stars at all the various epochs of those catalogs is made easy by the prior knowledge of their proper motion vectors.  

In addition to image-differenced proper motions, SUPERBLINK also incorporates positions and proper motions from Gaia DR1 where available \citep[][]{Gaia2016a, Gaia2016b}.  The Gaia proper motions are labelled {\tt \_new} in the SUPERBLINK catalog and the CDC.

As a base catalog for determining cool dwarfs for TESS observations, we therefore adopted the SUPERBLINK catalog as a primary source of targets.  Although SUPERBLINK already lists 2MASS magnitudes for most objects, we independently cross-matched the catalog again with 2MASS in order to obtain the 2MASS photometric flags.

SUPERBLINK contains Tycho-2 V-band magnitudes where available \citep[][]{Hog2000}.  More recent visible magnitudes are available from the American Association Variable Star Observers Photometric All-Sky Survey \citep[APASS,][]{Henden2014} catalog.  APASS contains Landolt B and V magnitudes, and SDSS-like $g'$, $r'$ and $i'$ magnitudes for roughly 2.5 million stars.  We therefore cross-matched SUPERBLINK with APASS in addition to 2MASS.

Cross-matching between the three catalogs was performed using a PostgreSQL database maintained at Vanderbilt University for the purpose of determining the TESS Input Catalog and the Candidate Target List \citep[TIC and CTL,][]{Stassun2017}.  The SQL database contains complete catalogs, which we cross-matched and filtered using an SQL query.  The resulting cross-matched catalog contained 2,704,792 entries.

\subsection{Determining a Visible-band Magnitude}\label{sec:vmag}

\begin{figure}
    \centering
    \includegraphics[width=\linewidth]{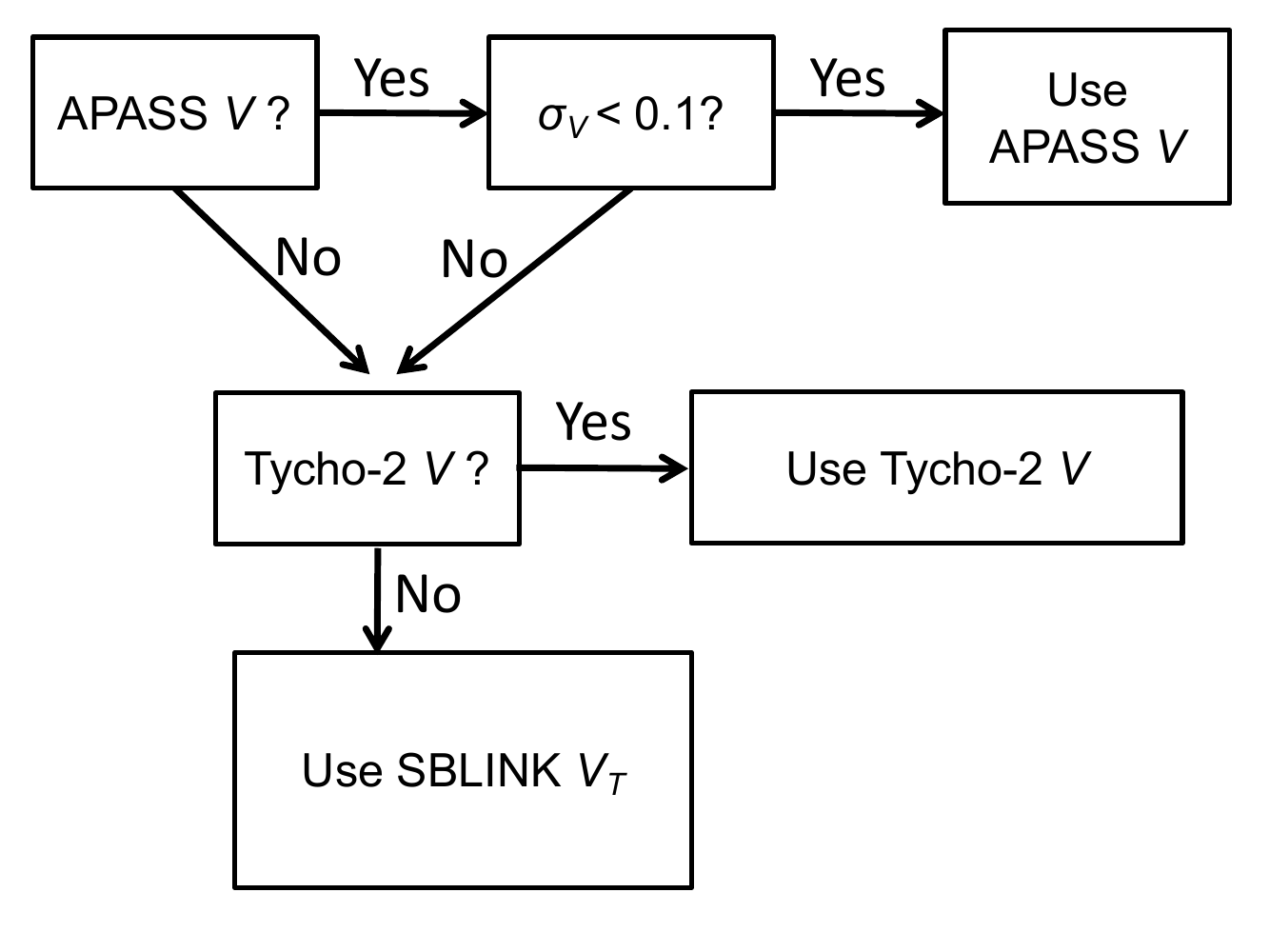}
    \caption{Decision chart indicating which archival measurement is used to determine a V-band magnitude for a given object in SUPERBLINK.}
    \label{v_mag_flowdown}
\end{figure}

As stated in the previous section, at least one visible-band magnitude provides a significant constraint on M dwarf effective temperatures.  We chose to use Johnson $V$-band owing to its wide availability.  For the SUPERBLINK/2MASS/APASS cross-matched catalog, we determined the best $V$-band magnitude based on the perceived reliability of APASS, Tycho-2 and the $V$-magnitude estimated in SUPERBLINK, called $V_T$.  If APASS V was available with 1$\sigma$ uncertainties of less than 0.1 magnitudes, we chose APASS $V$ as the $V$-band magnitude.  If, however, APASS was not available or had 1$\sigma$ uncertainties larger than 0.1 magnitudes, then we chose the Tycho-2 magnitude and assigned an uncertainty of either 0.013 or 0.1 magnitudes, for $V<9.0$ and $V \ge 9.0$ respectively, following \citet{Hog2000}.  If Tycho-2 magnitudes were not available, we chose the SUPERBLINK $V_T$ magnitude, an estimate of the star's $V$-band magnitude based on scanned plates.  The $V_T$ magnitude is described in \citet[][]{Lepine2011} and includes an estimate of the uncertainty.  Figure \ref{v_mag_flowdown} describes the process of choosing a $V$-band magnitude for each star.

With this V-band magnitude in-hand, we applied a criteria to isolate cool stars from stars with sun-like and hotter photospheric temperatures.  As a primary criterion for identifying cool dwarfs, we selected only stars with $V-J>2.7$, as suggested in \citet[][]{Lepine2011}. As demonstrated in \citet[][]{Lepine2013}, this selection should include nearly all dwarfs of subtype M0 and later, with only some contamination from late-type K dwarfs.  However, unlike \citet{Lepine2011}, we do not apply a brightness cut of $J<10$, in order to increase the number of stars in the catalog to a value closer to that predicted in S15, at the expense of perhaps including a few M giant contaminants to the target sample.

\subsection{Dwarf/Giant Separation}

To separate cool dwarf stars from more-massive, evolved stars with similar $V-J$ color, we used archival trigonometric parallax measurements, when parallaxes were available, and reduced proper motions where parallaxes are not available.  For stars with archival parallax observations, we applied the following selection criteria to isolate cool dwarfs, following the approach of \citet[][]{Gaidos2014}:

\begin{equation}
M_V > 2.2 \times (V - J) - 2.0 
\label{parallax_equation}
\end{equation}

Only 3535 stars in the combined catalog met this criterion.  For stars without archival trigonometric parallaxes, we used the reduced proper motion $H_V$:

\begin{equation}
H_V = V + 5.0 \times \log(\mu) + 5.0
\end{equation}

\noindent where $\mu$ is the proper motion in arc seconds per year.  Reduced proper motions have been used by previous authors to identify M dwarfs for exoplanet surveys \citep[e.g.][]{Lepine2011}.  Specifically, we adopted the reduced proper motion criteria of \citet[][]{Gaidos2014} for cool dwarfs, except for those with especially red colors.  \citet[][]{Gaidos2014} curated a list of bright ($J<10$) M dwarf stars for the purpose of identifying the best targets for radial velocity surveys.  In this work, we extend this approach to the full SUPERBLINK catalog in order to include fainter cool dwarfs, as suited for TESS:

\begin{equation}
\begin{split}
H_V > & \, 8.806 \, + \\ & 2.304 \times (V - J - 2.7) \, + \\ & 0.054 \times (V - J - 2.7)^2
\end{split}
\end{equation}

\begin{figure}
    \centering
 \includegraphics[width=\linewidth]{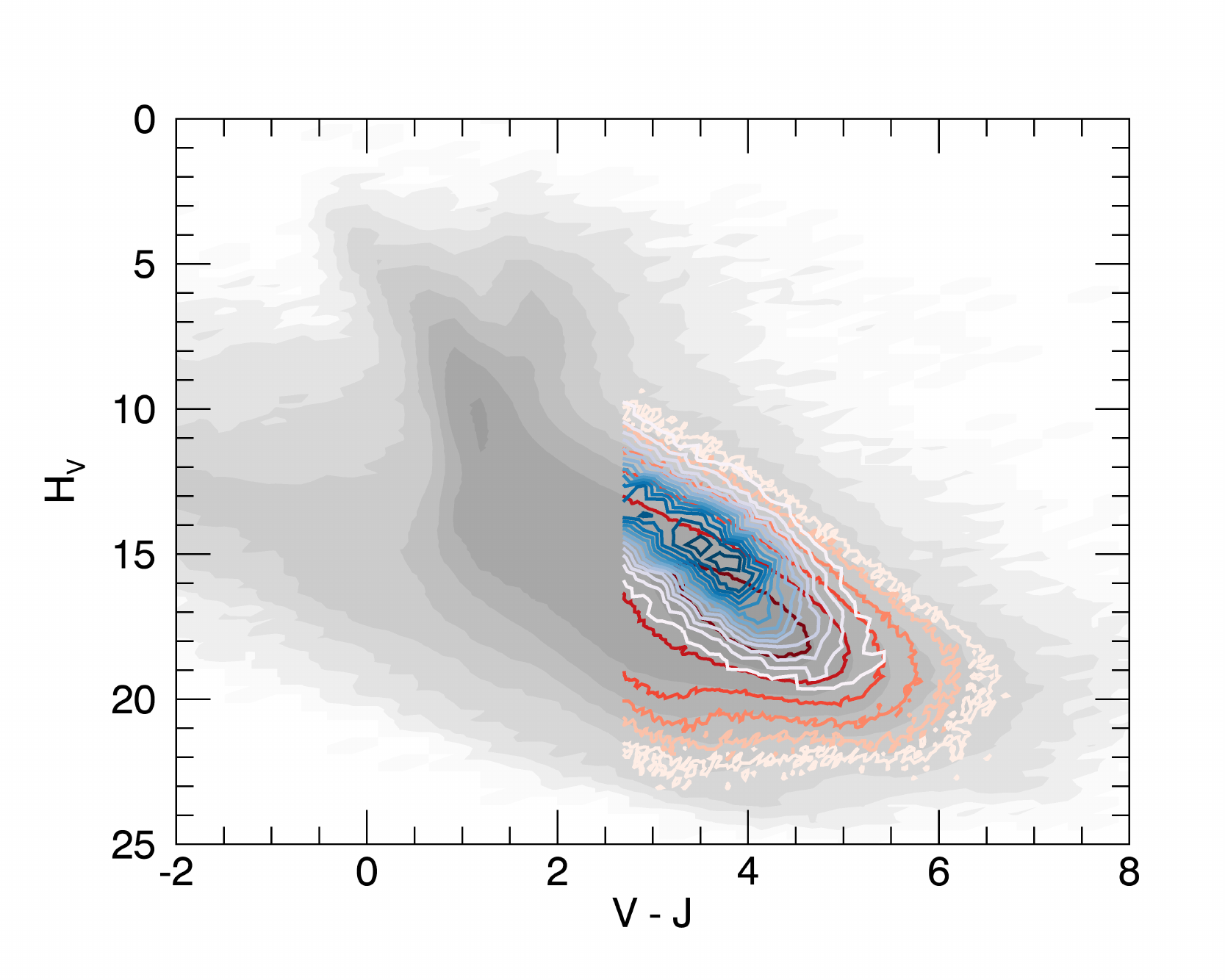}
 \caption{Contours showing the density of stars with reduced proper motion $H_V$ versus color $V-J$ for all stars in SUPERBLINK (white-to-black color scale), for dwarfs identified by reduced proper motion (white-to-red color scale) and dwarfs identified by trigonometric parallax (white-to-blue color scale).  Contours are logarithmically spaced.\label{rpm_fig}}
\end{figure}

Following the approach of \citet[][]{Lepine2011}, we further applied color criteria to help remove evolved stars that may otherwise pass the reduced proper motion criteria. \citet{Lepine2011} initially set relatively stringent color limits in [J-H,H-K] space, but these limits have since been found to exclude very late-type M dwarf stars such as TRAPPIST-1; hence we have modified the boundaries to include such objects. For stars with $H-K<0.25$, we followed the color criteria of \citet[][]{Lepine2011}:

\begin{equation}
J - H < 1.0 \\
\end{equation}
\begin{equation}
0.746 < J - K_S < 0.914    
\end{equation}

However for stars with $H-K \ge 0.25$ we adopted a more liberal limit.  This was invoked to increase the numbers of mid-to-late M dwarfs that may have been excluded by the criteria of \citet[][]{Lepine2011}:

\begin{equation}
J - H < 0.914    
\end{equation}

\begin{figure}
    \centering
 \includegraphics[width=\linewidth]{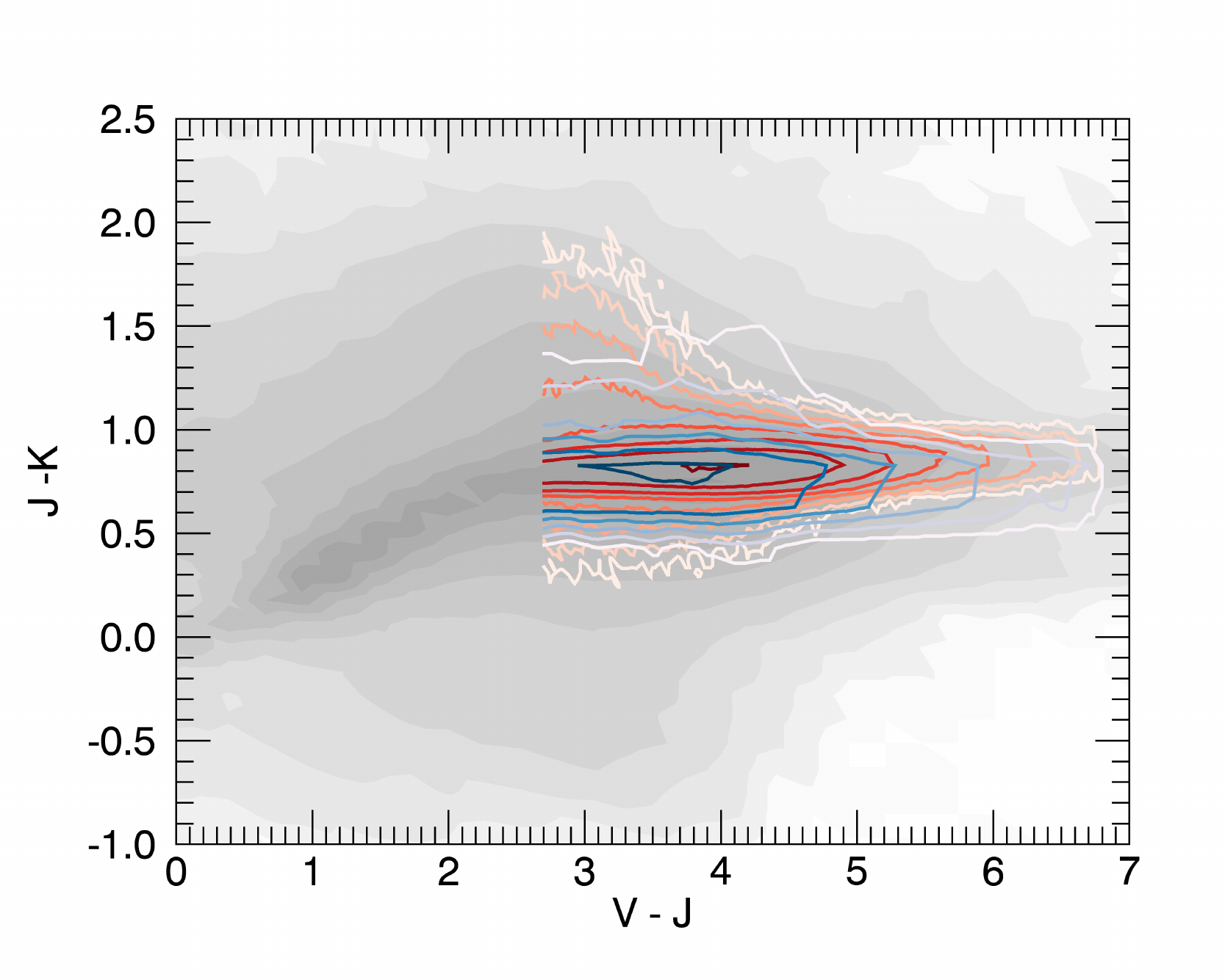}
 \caption{Contours showing the density of stars with colors $J-K$ versus $V-J$.  The color coding is identical to Figure \ref{rpm_fig}.}
 \label{vj_jk_fig}
\end{figure}

\begin{figure}
    \centering
 \includegraphics[width=\linewidth]{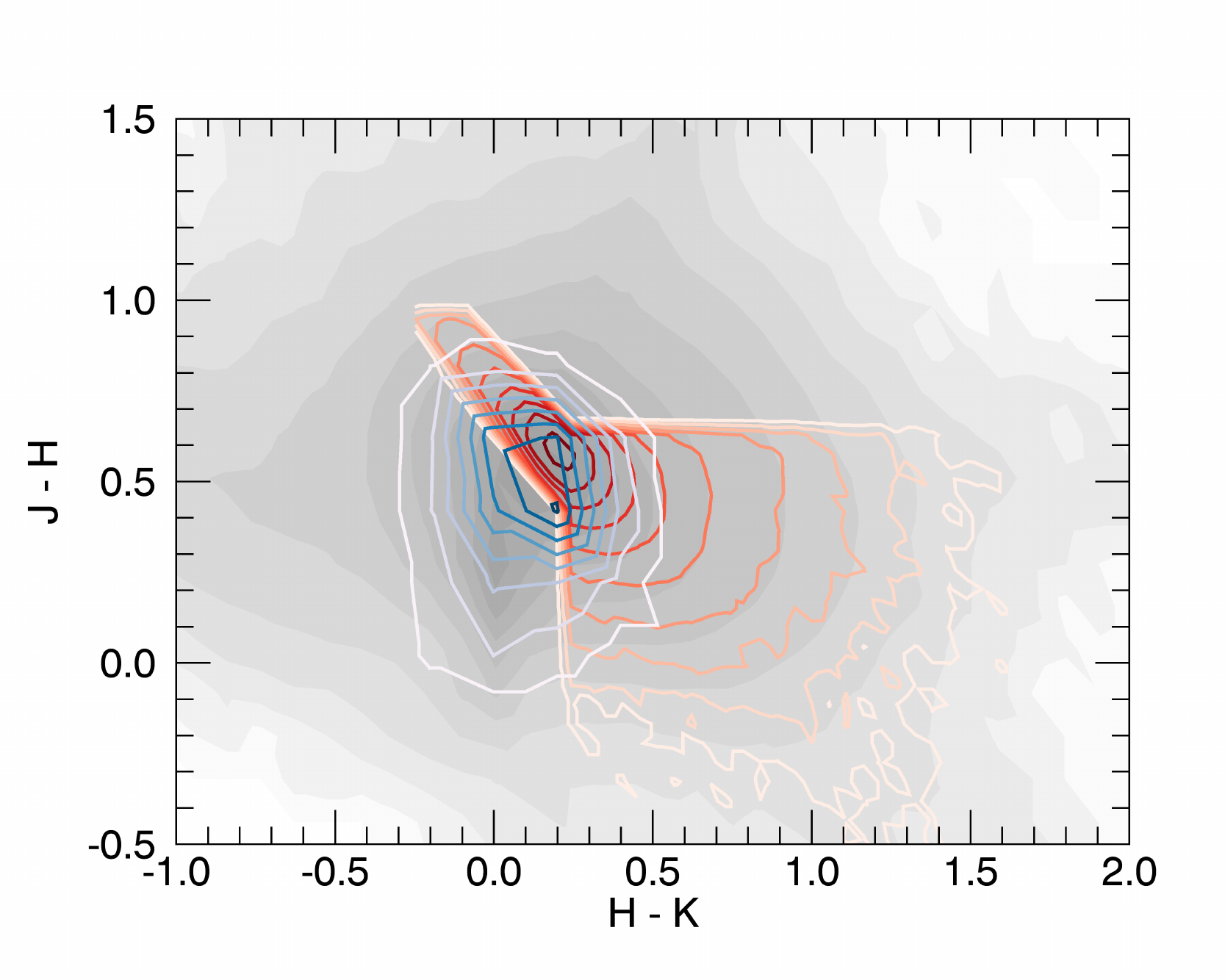}
 \caption{Contours showing the density of stars with colors $J-H$ versus $H-K$.  The color coding is identical to Figure \ref{rpm_fig}.}
 \label{hk_jh_fig}
\end{figure}

Figures \ref{rpm_fig}, \ref{vj_jk_fig} and \ref{hk_jh_fig} show all stars in the SUPERBLINK catalog and the stars identified as cool dwarfs using these criteria, separated into the parallax and proper motion identified stars.  Because SUPERBLINK contains stars with high proper motions, giants are largely excluded in the catalog to begin with.  After applying these criteria to the combined catalog, we are left with 1,076,470 stars that meet the selection criteria, in addition to those with archival trigonometric parallax observations.

\begin{figure*}
    \centering
    \includegraphics[width=\linewidth]{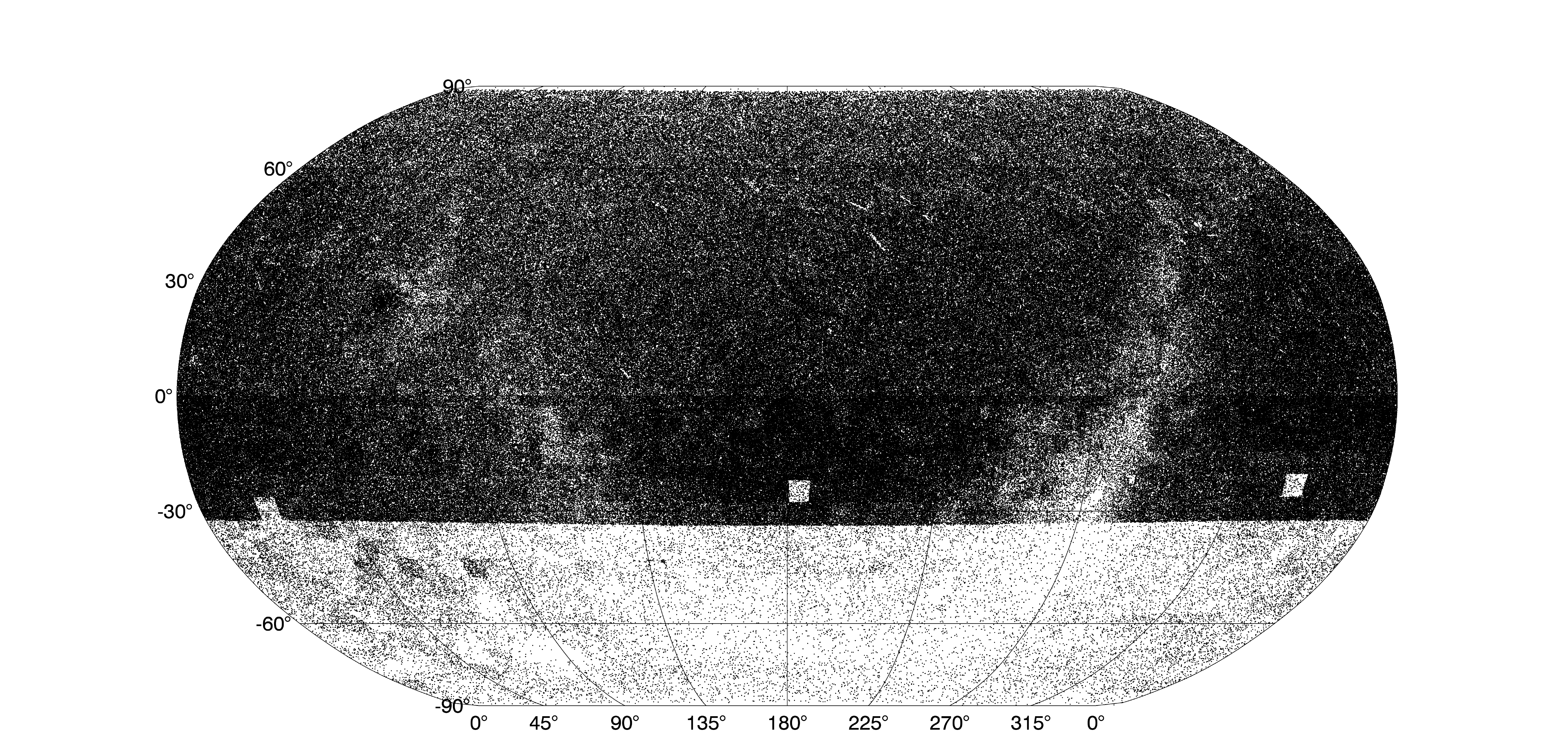}
    \includegraphics[width=\linewidth]{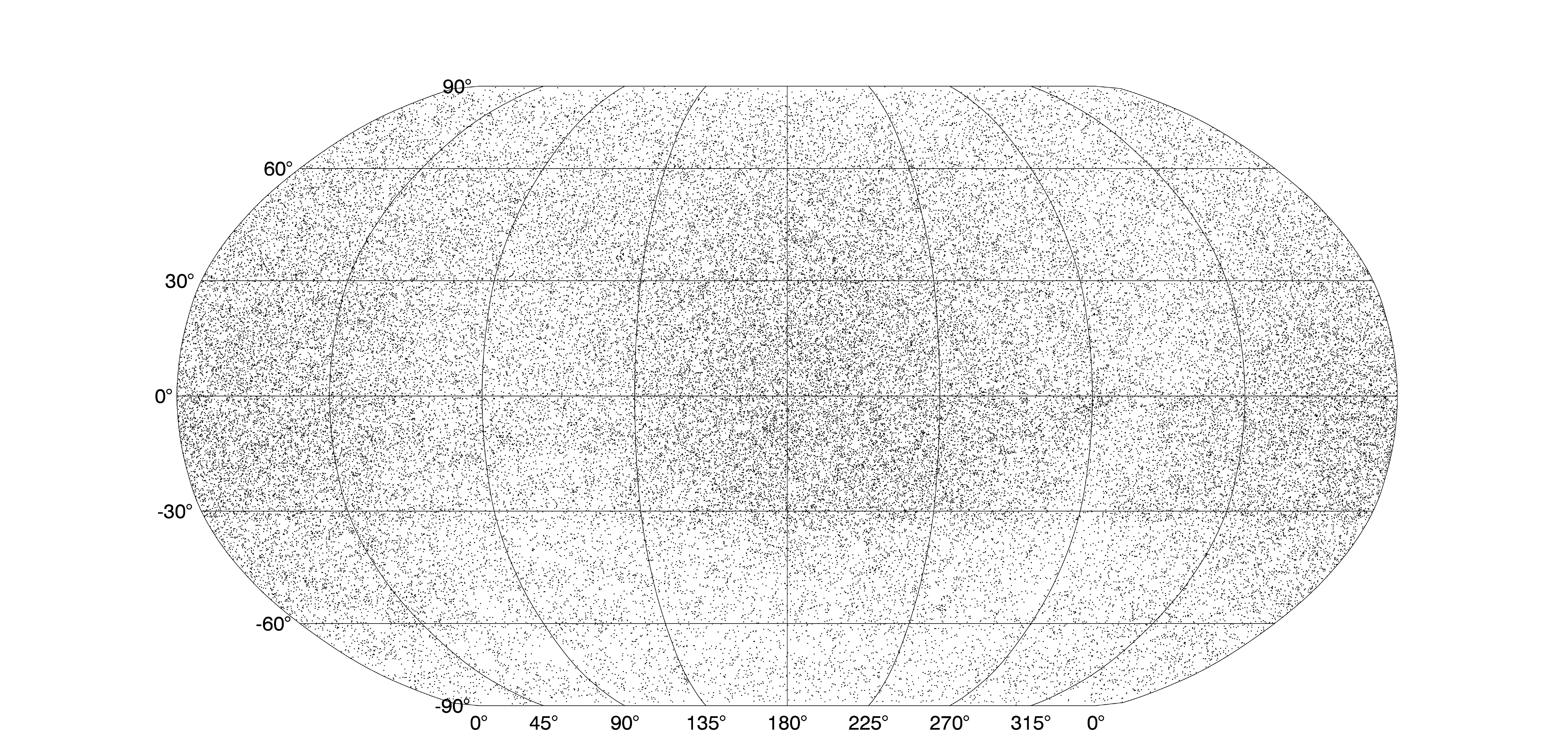}
    \caption{{\it Top:} Projected sky image of the locations of stars in the cool dwarf catalog in equatorial coordinates.  The lack of cool dwarfs with $\delta < -32^\circ$ is a consequence of a lack of long-time-baseline proper motion data incorporated into SUPERBLINK. {\it Bottom}: Same as the top image, but only including stars with proper motions greater than 150 {\it mas}, where the catalog is more complete in the south.}
    \label{catalog_radec}
\end{figure*}

SUPERBLINK is known to be less complete for declinations ($\delta$) less than $-30^\circ$.  This is due to the coverage of POSS I and POSS II, which could not observe stars south of $\delta=-30^\circ$ due to the latitude of the 48-inch Samuel Oschin Telescope at Palomar Observatory.   Figure \ref{catalog_radec} shows the resulting catalog of stars across the night sky, in equatorial coordinates.  

There are 1,039,173 stars in the cool dwarf catalog with $\delta \ge -30^\circ$ and only 40,832 with $\delta < -32^\circ$,  owing to the lack of 1950s plate data from POSS-I.  However, if only proper motions of greater than 150 {\it mas} are considered, the spread of objects is actually uniform across the sky.  The significant lack of stars with $\delta<-32^\circ$ presents a challenge to realizing the star counts in S15.   Three patches of sky above $-32^\circ$ represent three fields for which 1950s plate data (from POSS-I) is unavailable from the Digitized Sky Surveys.

\subsection{Comparison to Galactic Simulations}

To explore the completeness of the resulting catalog of cool dwarfs, we can compare the star counts to those predicted by galactic simulations.  To simulate the M dwarf counts expected in the night sky, we used the TRIdimensional modeL of thE GALaxy \citep[TRILEGAL, pronounced TREE-leh-GOW,][]{Girardi2005}, and followed an approach nearly identical to S15.  We used the HEALPix software to determine 3072 equally spaced coordinates across the sky, each centered on a region subtending a solid angle of 13.4 deg$^2$ \citep{Gorski2005}.  We neglected coordinates with galactic latitudes less than 8$^\circ$ due to their proximity to the galactic plane, where TRILEGAL computations take excessively long.  For each coordinate, we queried the TRILEGAL web interface\footnote{http://stev.oapd.inaf.it/cgi-bin/trilegal} using a Perl script written by L. Girardi and available as part of the VESPA software package \citep[][]{Morton2012, Morton2015}.

We used the default options in TRILEGAL, but with a solid angle of 6.7 deg$^2$, half the 13.4 deg$^2$ corresponding to each coordinate in order to decrease the computation time.  To account for this discrepancy, each star in the resulting simulation is doubled, replicating the approach by S15.  We applied a magnitude cut of $V<20$ as the cool dwarf catalog does not contain stars fainter than this limit.  Once downloaded, we applied an absolute magnitude cut to the simulated stars matching the cut used for cool dwarfs with parallax observations (Eq. \ref{parallax_equation}). 

\begin{figure}
    \centering
    \includegraphics[width=\linewidth]{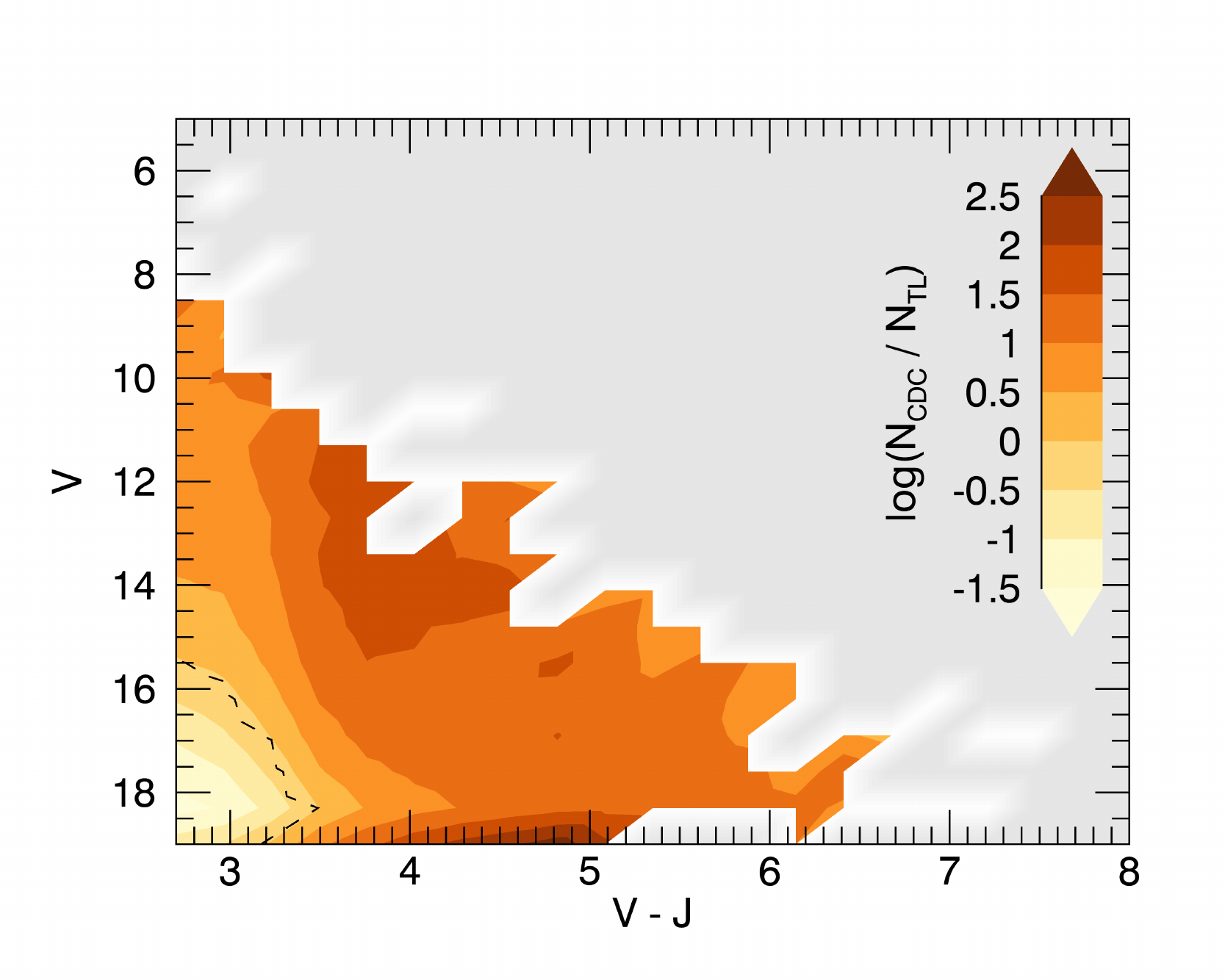}
    \caption{Contour plot showing the log of the number of stars in the cool dwarf catalog (CDC) divided by the number of stars returned by TRILEGAL, versus $V$ magnitude and $V-J$ color, for $\delta>-30^\circ$.  The contours depict an estimate of the completeness of the catalog compared to models of the galaxy.  Grayed areas contain no stars in TRILEGAL, and the dashed contour corresponds to equal numbers of stars in both.}
    \label{compare_to_trilegal_vj_north}
\end{figure}

\begin{figure}
    \centering
    \includegraphics[width=\linewidth]{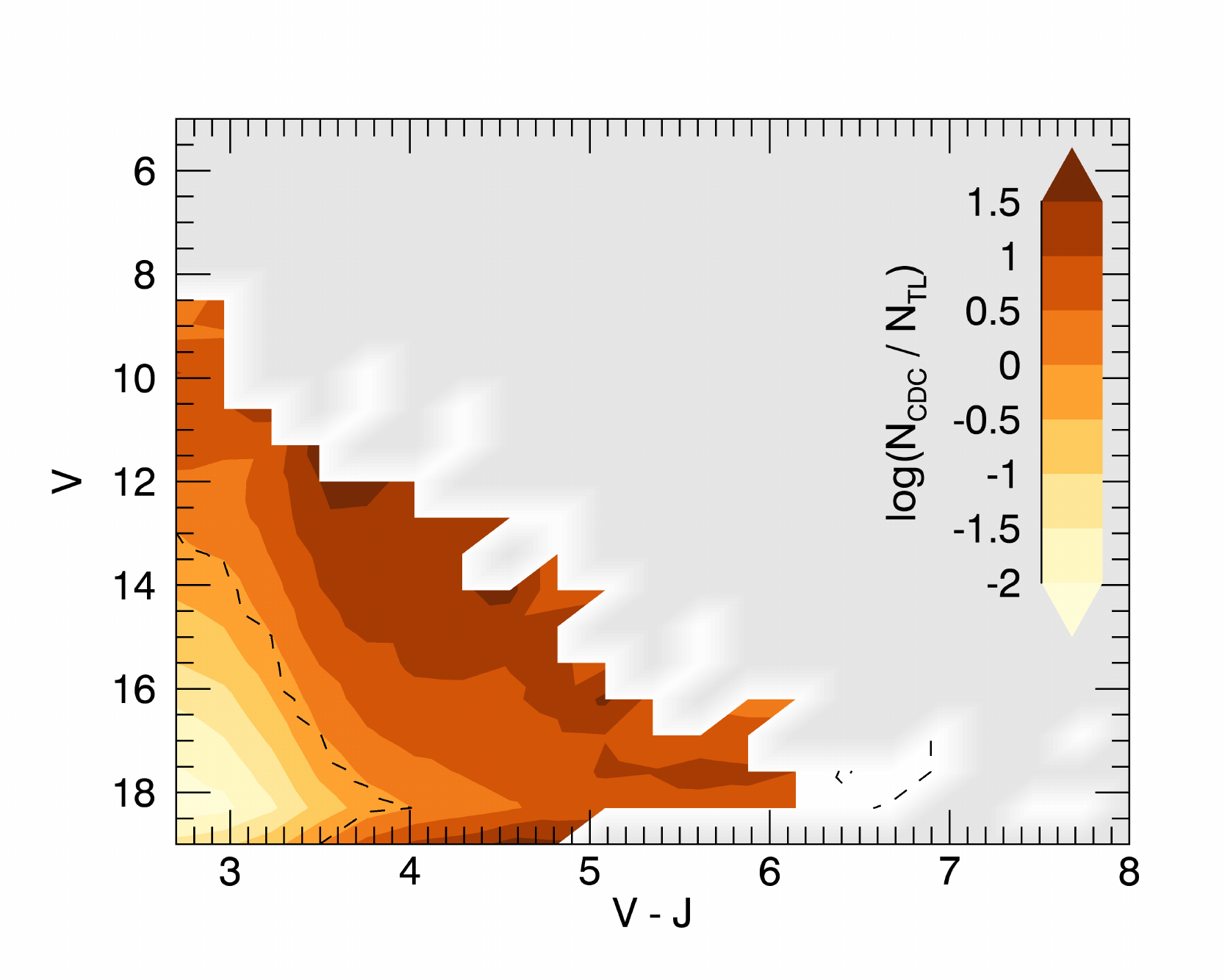}
    \caption{Same as Figure \ref{compare_to_trilegal_vj_north}, but for $\delta<-30^\circ$.}
    \label{compare_to_trilegal_vj_south}
\end{figure}

We compared the simulated stars from each TRILEGAL pointing to the detected stars in the cool dwarf catalog, within that same solid angle on the sky.  Due to the incompleteness of SUPERBLINK for $\delta<-30^\circ$ we divided the comparison into 'north' and 'south' groups corresponding to that boundary.  Figure \ref{compare_to_trilegal_vj_north} shows the resulting completeness of the cool dwarf catalog as a function of $V$ and $V-J$ for stars with $\delta>-30^\circ$ and Figure \ref{compare_to_trilegal_vj_south} shows the same for stars with $\delta<-30^\circ$.  In both cases, the cool dwarf catalog has more supposed dwarf stars than predicted for most of the phase space, indicating potential contamination by giant stars with erroneous proper motion measurements.  This is especially concerning near $V-J\sim4$ and $V\sim13$, where the cool dwarf catalog has over 100 times as many objects as predicted by TRILEGAL in the 'north' sample.  We suspect the inaccuracies of the magnitudes are contributing significantly to the discrepancies in star counts between the catalog and predictions from TRILEGAL.

\subsection{Interstellar Extinction and Reddening}

Interstellar extinction and reddening can bias the catalog selection, especially for distant stars, where these effects can be especially strong.  However, the cool dwarfs considered here are not particularly distant, meaning extinction and reddening may not significantly affect the goals of the catalog.  As stated earlier, the optical minus infrared color provides the most leverage when determining an M dwarf temperature.  The color chosen in this work, $V$-$J$, may be affected by reddening. To assess the expected degree of extinction and reddening, we examined the values for extinction in the $V$-band ($A_V$) listed in the TRILEGAL simulations.  The so-called standard interstellar extinction law measured by \citet{Rieke1985} calculates $A_J/A_V$ of 0.282 for ``standard'' lines of sight.  Combining these two, we can determine the largest effect expected from interstellar extinction and reddening.

\begin{figure}
    \centering
    \includegraphics[width=\linewidth]{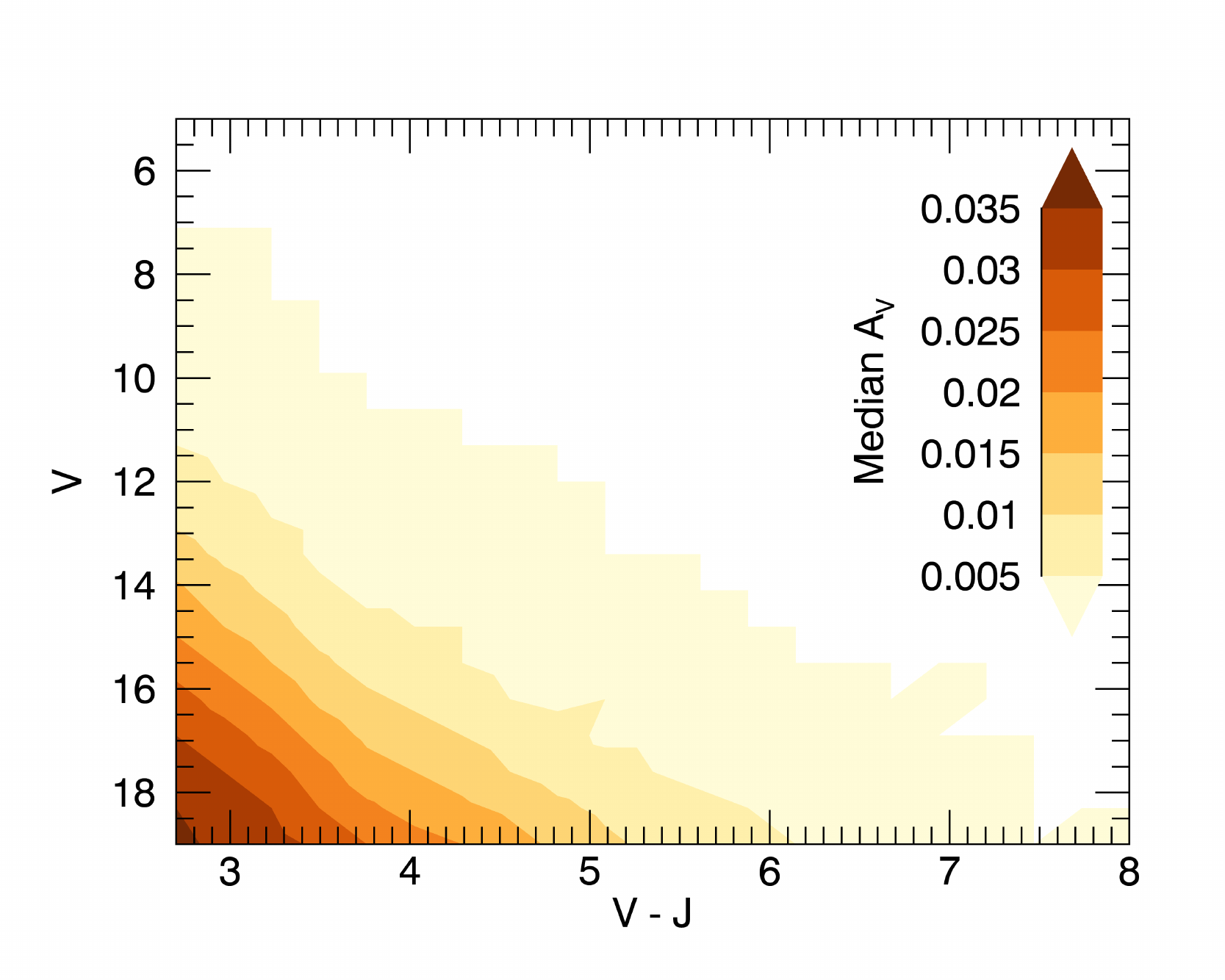}
    \caption{Contour plot showing the median $A_V$ in the TRILEGAL simulations with dwarf criteria applied, versus $V$ magnitude and $V-J$ color.  The maximum extinction is expected to be $A_V=0.038$, and that primarily applies to the faintest stars and bluest objects in the cool dwarf catalog, which by nature are the most distant.}
    \label{trilegal_extinction}
\end{figure}

Figure \ref{trilegal_extinction} shows the median extinction $A_V$ versus $V$ magnitude and $V-J$ color for the TRILEGAL simulations, subjected to the dwarf selection criteria described previously.  As expected, $A_V$ is largest for the faintest and bluest stars, peaking at a value of 0.038.  Combining this with $A_J/A_V$ measured by \citet{Rieke1985}, dwarf stars in the cool dwarf catalog are not expected to be reddened by more than $\Delta(V-J)=0.027$.  Given the relatively small size of the effect, we choose to ignore extinction and reddening in our criteria for determining the cool dwarf catalog, as well as for estimating stellar properties, for the sake of reproducibility.  We note, however, that any evolved stars that may contaminate the cool dwarf catalog are likely distant and significantly reddened.

\subsection{Binarity}

Stellar multiplicity of the target sample will have a significant impact on the scientific goals of the TESS Mission.  For widely separated multiple star systems, different apertures can be used to search for transiting planets orbiting each component. However, these objects may not be the best use of limited two-minute observations by TESS.  Recent work by \citet[][]{Kraus2016} showed that stars in multiple systems are less likely to host exoplanets, presumably due to gravitational effects on the individual stars' protoplanetary disks, or subsequent evolution of planetary orbits.  

Unresolved binary or multiple systems present their own challenges.  In the case of an unresolved equal mass binary, a given transiting planet will produce a shallower transit depth.  This might lead one to abandon any stars that appear to be short-period and unresolved binaries.  At the same time, two-minute observations of eclipsing binary stars have proven useful for determining highly precise and accurate parameters for the component stars, which are useful for determining the properties of planets found to orbit other single stars.  Recently, \citet[][]{Shan2015} calculated the ``number of M dwarfs per M dwarf'', in an attempt to determine the short-period binary rate among M dwarf stars using {\it Kepler} eclipsing binary data. They found 0.11$^{+0.04}_{-0.04}$ of presumably single M dwarfs are actually M dwarfs with a  short-period ($<90 day$) M dwarf companion. However, we note that \citet{Fischer1992} found a significantly higher overall M dwarf binary fraction of 42 +/- 9 \%, but for orbital periods out to 30 years.

Regardless of whether multiple stars systems, resolved or unresolved, should or should not be included in the cool dwarf target list, the fact is that without comprehensive spectroscopy, high-resolution spacial imaging, or trigonometric parallax data, determining multiplicity is prohibitively difficult.  The vast majority of stars identified as cool dwarf stars using the approach described above do not have sufficient data to determine whether they host an equal mass or lower mass binary star, or multiple stars.  For these reasons, we choose to ignore binarity in the determination of this cool dwarf target list. In the proceeding analysis we treat every object in the cool dwarf catalog as a single star.  For the vast majority of targets in the Cool Dwarf Catalog, this will have little affect on the errors in the estimated stellar parameters, since the colors are dominated by the primary component.  Ironically, this assumption will lead to the largest error in the parameters for stars with trigonometric parallax measurements, whereby the absolute magnitude of a binary is presumed to be due to a single star.  

\subsection{TESS Magnitudes}

In order to choose the best cool dwarfs from the list for two-minute observations, we must determine the brightness of the stars within the TESS band, or equivalently the TESS magnitude $T$.  The TESS response function is described in \citet{Ricker2015}.  Briefly, the response function is optimized for the red-end of the visible spectrum, rising sharply redward of 600 nm, then descending gradually from 900 to 1050 nm.  As described in S15, in a Vega system $T$ is most similar to Cousins $I$-band magnitude, or $I_C$.  Unfortunately, $I_C$ is largely unavailable for the vast majority of stars in the catalog.  SDSS-like $i'$-band is available for many stars via APASS; otherwise, $V$, $J$, $H$ and/or $K_s$ can be used to determine the TESS magnitude, as long as effects from spectral type are taken into account.

To assess the role of spectral type when converting magnitude(s) to $T$, we calculated synthetic photometry from 183 spectro-photometrically calibrated spectra of nearby M dwarf stars. \citet[][]{Mann2015} stitched together visible and near-infrared spectra of nearby M dwarf stars and photometrically calibrated the resulting full spectra.  Being stitched across multiple bands and spectrophotometrically calibrated, the spectra are useful for determining conversions between various visible and infrared bands as a function of color or spectral type.  The 183 calibrator stars span spectral types from K7 to M7, temperatures from 2700 to 4100 K, masses from 0.08 to 0.74 $M_\sun$ and iron abundances ([Fe/H]) from $-0.61$ to 0.53.

For each of the spectro-photometrically calibrated spectra, we multiplied by the response functions for $i'$, $T$ and for 2MASS $J$, $H$ and $K_s$.  For each case, we calculated the resulting colors and fitted three magnitude conversion equations: one using $i'$ and $K_s$ magnitude, and one using $V$ and $K_s$ and one using $J$, $H$ and $K_s$:

\begin{equation}
\label{tik}
T =  0.6262 + 1.0004 \times i' - 0.3326 \times (i' - K_S)
\end{equation}

\begin{equation}
\label{tvk}
T = 0.8536 + 0.9972 \times V - 0.6793 \times (V - K_S)
\end{equation}

\begin{equation}
\label{tjhk}
\begin{split}
T = 0.6316 & + 1.0329 \times J \\ & - 0.4324 \times (J - H) \\ & + 4.0239 \times (H - K_S)
\end{split}
\end{equation}

The root-mean-square residuals for each of the fitted relations is 0.030, 0.055 and 0.14 magnitudes, respectively.  The final relation has the largest scatter because it uses infrared magnitudes to determine a visible magnitude.  However, for many of the stars in the cool dwarf catalog, only 2MASS magnitudes are uniformly measured.

\begin{figure}[htbp]
    \centering
    \includegraphics[width=\linewidth]{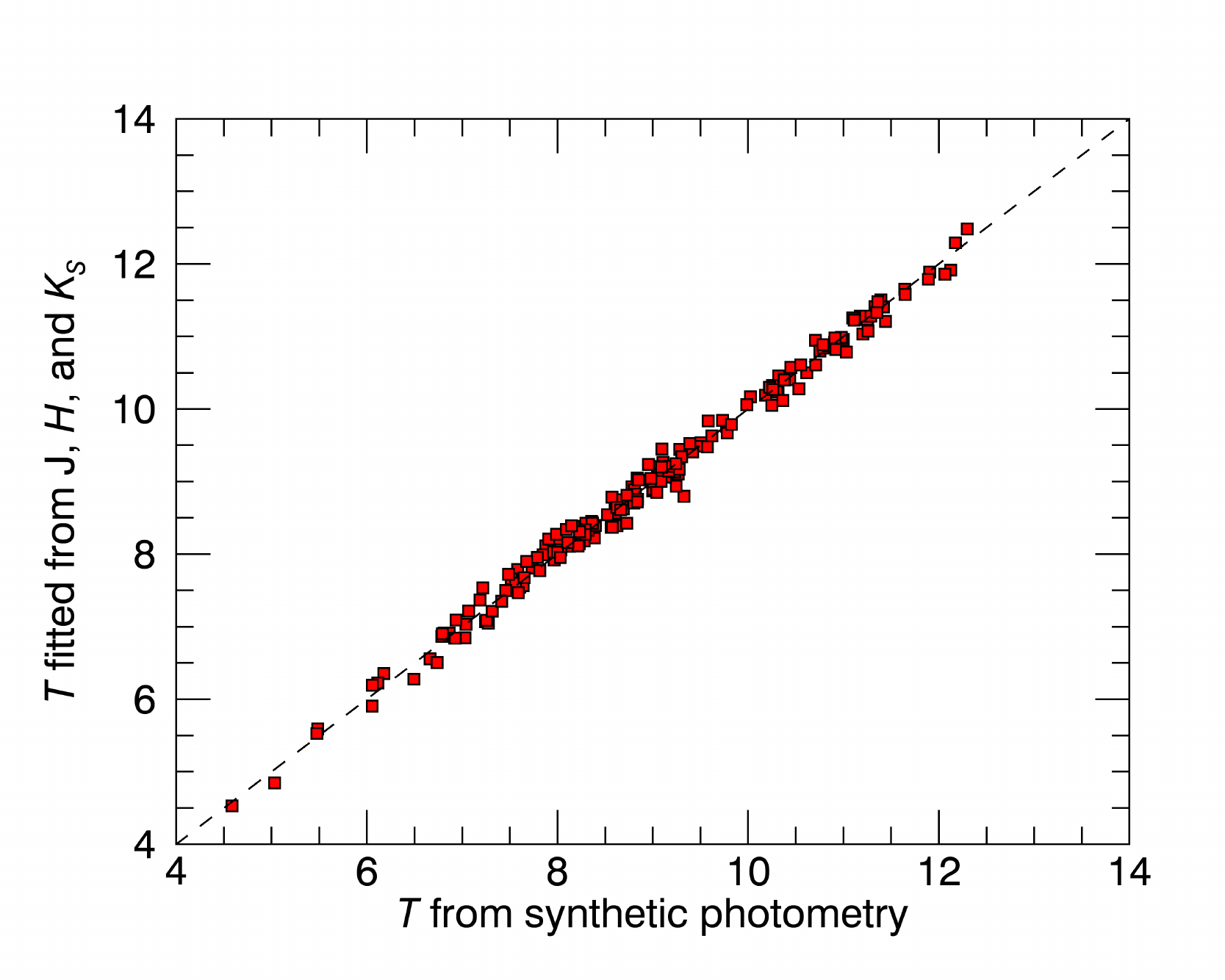}
    \caption{Synthetic TESS magnitude ($T$) versus $T$ derived from the conversion from 2MASS magnitudes (see Equation \ref{tjhk}) for spectro-photometrically calibrated spectra from \citet{Mann2015}.  The root-mean-square of the residuals is 0.14 magnitudes.}
    \label{tess_mag_plot}
\end{figure}

Figure \ref{tess_mag_plot} plots $T$ calculated via synthetic photometry versus $T$ determined using Equation \ref{tjhk}, for the calibration spectra.  To determine $T$ for each star in the cool dwarf catalog, we use Equation \ref{tik} if APASS or SDSS $i'$ is available.  However, we found that for the calibration sample, the SDSS synthetic magnitudes did not match measured APASS magnitudes of the stars.  We applied a conversion from APASS $i'$ to SDSS $i'$ prior to using Equation \ref{tik} based on this discrepancy.

If $i'$ is not available, we used Equations \ref{tvk} and \ref{tjhk} and take the weighted mean between the two, weighting by their respective uncertainties.  For those uncertainties, we propagate the reported or assigned $V$, $i'$, $J$, $H$ and $K_s$ uncertainties through the equations and in the case of Equation \ref{tjhk}, we add a systematic uncertainty of 0.14 magnitudes in quadrature to account for the systematic error in the fitted conversion.

\subsection{Effective Temperatures}
\label{ssec:teff}
We determine an effective temperature (\teff) for each cool dwarf in the catalog using the color-\teff\ relations from \citet[][]{Mann2015} and \citet{Mann2016}. Specifically, we use the $r-J$, $r-z$, and/or $V-J$ relations, depending on available magnitudes, and we include a $J-H$ term to account for systematic effects of metallicity on the resulting effective temperature \citep[see][]{Johnson2012,Mann2013a,Newton2014}. Because of the requirements for target selection (Section~\ref{sec:vmag}), all targets have at least one of these colors, with all having at least a $V$ and $J$ magnitude. For targets with more than one color available we use the weighted mean of the derived \teff\ as the final value.  We note that very few targets have reliable $z$-band observations, and that particular relation was rarely invoked.

For stars with APASS $r'$ magnitudes, we again applied a correction to convert to SDSS $r'$ magnitudes, based on the discrepancy between measured APASS $r'$ and synthetic SDSS $r'$ magnitudes on the spectro-photometric calibration sample.

Errors on \teff\ are computed by combining errors due to magnitude uncertainties, scatter in the color-\teff\ relations, and the intrinsic scatter in the \citet{Mann2015} calibration sample (60~K).  The calibration sample of \citet{Mann2015} contains no stars with $V-J>7$ ($r-J\lesssim6.2$, \teff\ $\lesssim$ 2700).  For stars beyond this limit we do not assign a \teff, but we keep those stars in the catalog for completeness.

\subsection{Masses and Radii}

For cool dwarfs with trigonometric parallaxes, we first calculate the absolute $K_s$ magnitude and associated error from its 2MASS magnitude. We convert $M_{K_s}$ to mass using the  $M_{K_s}$-$M_\star$ relation from \citet{Benedict2016}, a 5th degree polynomial.  This relation was calculated using dynamical masses of low-mass binary stars, and is valid for $5.2<M_{K_s}<11$.  To calculate the uncertainty in mass, we propagated the uncertainty $M_{K_s}$ through the polynomial, as well as the uncertainties in the polynomial coefficients themselves.  The coefficients have non-zero covariances, and we include the reported covariances when determining the uncertainty.

For cool dwarfs with $4.5<M_{K_s}<5.2$ we use the \citet{Delfosse2000} $M_{K_s}$-$M_\star$ relation, which was similarly calibrated on dynamical masses, but had fewer calibration stars than \citet{Benedict2016}.  However, the \citet{Benedict2016} relations are not valid over this regime whereas the \citet{Delfosse2000} are.  We follow an uncertainty procedure nearly identical to what is describe above, but do not propagate the uncertainties in the coefficients themselves, as these are not reported in \citet{Delfosse2000}.

We then calculate stellar radii using the $M_{K_s}$-$R_\star$ relation from \citet{Mann2015}, which is precise to $\simeq$3\% ignoring errors in $K_s$ and distance.  We calculated uncertainties in stellar radius by propagating the uncertainty in $M_{K_s}$ through the relationship, then incorporate a 3\% systematic uncertainty by adding the uncertainty in quadrature.

For stars lacking a trigonometric parallax, or the vast majority of the catalog, we determine radius using the \teff-$R_\star$ relation from \citet{Mann2015}. For masses we derive a new relation between \teff\ and $M_\star$ based on the 183 M dwarfs with precise distances and radii collected by \citet{Mann2015}, but deriving new masses for these stars based on the \citet{Benedict2016} $M_{K_s}$-$M_\star$ relation to keep the masses consistent with our parallax-based values.

Radii and masses derived from \teff\ are almost always less precise than those derived using $M_{K_s}$, with best case errors on radius of $\simeq$3\% in the latter case, and 13\% in the former. This is primarily due to the important role metallicity plays in the relation between luminosity and \teff\ for M dwarfs, of which $M_{K}$ is relatively immune \citep{Delfosse2000,Mann2015}, and the steep relation between \teff\ and $R_\star$ compared to between $M_{K_s}$ and $R_\star$ for M dwarfs. 

\subsection{The Cool Dwarf Catalog}

The final catalog contains 1,080,005 entries.  Table \ref{column_headings} shows the column headings for the catalog.  We report identifiers, the J2000 equatorial coordinates, the chosen $V$ magnitude, the calculated TESS magnitude $T$, $T_{\rm eff}$, radius, mass,  corresponding uncertainties and flags.  For each entry, we also include the columns available from SUPERBLINK, 2MASS and APASS verbatim for completeness.  If a particular parameter could not be computed, or is not present in a catalog, the specific entry is left blank.

The J2000 coordinates are taken from SUPERBLINK, which sets all the coordinates to the 2000.0 epoch, extrapolating from the 2MASS coordinates/epochs using the measured proper motion vectors.  The flag assigned to the $V$ magnitude can take one of three strings: {\tt apass}, {\tt tycho} or {\tt sblink}, indicating the source of the chosen $V$ magnitude.  The flag assigned to the TESS magnitude $T$ can take one of five flags: {\tt from\_apass\_ik}, {\tt from\_sdss\_ik}, {\tt wmean\_vk\_jhk}, {\tt vk} or {\tt no\_kmag}.  The first two use a combination of $i$ band (from either APASS or SDSS, respectively) and 2MASS $K$ band to determine $T$.  The third uses a weighted mean between the $V-K$ relation and the 2MASS only relation.  The fourth uses only $V-K$, and the fifth indicates that the object does not have a reliable $K$-band magnitude, so a TESS magnitude could not be reliably determined.

The \teff\ flag is a combination of letters indicating which colors were used to determine the effective temperature.  The combination can include {\tt rj}, {\tt vj}, and/or {\tt rz}, each corresponding to the color used.  The final \teff\ is the weighted average of all the color-\teff\ relations used.  The radius flag can take one of three strings: {\tt from\_teff}, indicating that it was determined from the \teff, {\tt from\_mk}, indicating it was determined from the absolute $K$-band magnitude \citep[][]{Mann2015}, or {\tt no\_radius}, indicating that it could not be determined.  Similarly, the mass flag can take one of several strings: {\tt from\_teff}, {\tt from\_mk\_D00}, indicating the \citet[][]{Delfosse2000} relations were used, {\tt from\_mk\_B16}, indicating the \citet[][]{Benedict2016} relations were used, or {\tt no\_mass}, indicating it could not be determined.

\begin{table}[]
    \centering
    \begin{tabular}{l l}
    \hline
    \hline
    { Column Heading} & { Description} \\
    \hline
    \hline
    tmcntr & 2MASS Identifier \\ 
    sblink & SUPERBLINK identifier \\
    ra & Right ascension (J2000)\\
    dec & Declination (J2000) \\
    v mag & $V$ magnitude \\
    v mag err & 1$\sigma$ uncertainty in $V$ \\
    v mag flag & Flag indicating $V$ source\\
    tess mag & $T$ (TESS) magnitude\\
    tess mag err & 1$\sigma$ uncertainty in $T$ \\
    tess mag flag & Flag indicating $T$ source\\
    teff & Assigned $T_{\rm Eff}$ (K)\\
    teff err & 1$\sigma$ Uncertainty in $T_{\rm Eff}$ (K)\\
    teff flag & Flag indicating $T_{\rm Eff}$ source\\
    radius & $R_\star$ ($R_\Sun$)\\
    radius err & 1$\sigma$ Uncertainty in $R_\star$ ($R_\Sun$)\\
    radius flag & Flag indicating $R_\star$ source \\
    mass & $M_\star$ ($M_\Sun$)\\
    mass err & 1$\sigma$ Uncertainty in $M_\star$ ($M_\Sun$)\\
    mass flag & Flag indicating $M_\star$ source \\
    sblink\_ & columns verbatim \\
     & from SUPERBLINK \\
    apass\_ & columns verbatim \\
     & from APASS DR9 \\
    tmass\_ & columns verbatim \\
     & from 2MASS \\
    \hline
    \end{tabular}
    \caption{Column Headings for the Cool Dwarf Catalog}
    \label{column_headings}
\end{table}

\subsection{Incorporation into the TIC and CTL}

The Cool Dwarf Catalog is currently incorporated into the TESS Input Catalog (TIC) and TESS Candidate Target List (CTL), both of which are described in \citet[][]{Stassun2017}.  Briefly, the TIC is an omnibus catalog with the goal of listing all stars that fall within the field of view and brightness limit of TESS during the primary mission, similar in scope to the {\it Kepler} Input Catalog for the {\it Kepler} Mission \citep[][]{Batalha2010, Brown2011}.  The CTL is a subset of the TIC containing stars that will be observed using 2-minute cadence, specifically for searching for transiting exoplanets.  

The TIC master database stores the Cool Dwarf Catalog in its own table.  The Cool Dwarf Catalog is matched against the rest of the TIC entries.  In the event that stars in the Cool Dwarf Catalog already exist in the TIC, the TIC retains both parameters (those already in the TIC and those in the Cool Dwarf Catalog); however, when choosing stars for the CTL, the process uses the parameters in the Cool Dwarf Catalog, as determined by the methods described in this paper.  When revised versions of the Cool Dwarf Catalog have been uploaded to the TIC, the TIC retains the superseded versions of the Cool Dwarf Catalog, but the CTL does not.  When revised versions of the Cool Dwarf Catalog are submitted, stars deleted from the prior Cool Dwarf Catalog are deleted from the CTL.


For TESS, the Cool Dwarf Catalog serves as an ``overriding, curated catalog.'' Stellar parameters such as TESS magnitude, radius, effective temperature and derived values already in the TIC are overwritten by values reported in the Cool Dwarf Catalog. There are a few cases, however, where effective temperature and/or radius are not been reported in the Cool Dwarf Catalog. In these cases the values already in the TIC are retained.

\subsection{Comparisons between Parallax and Magnitude Determinations}

\begin{figure}
    \centering
    \includegraphics[width=\linewidth]{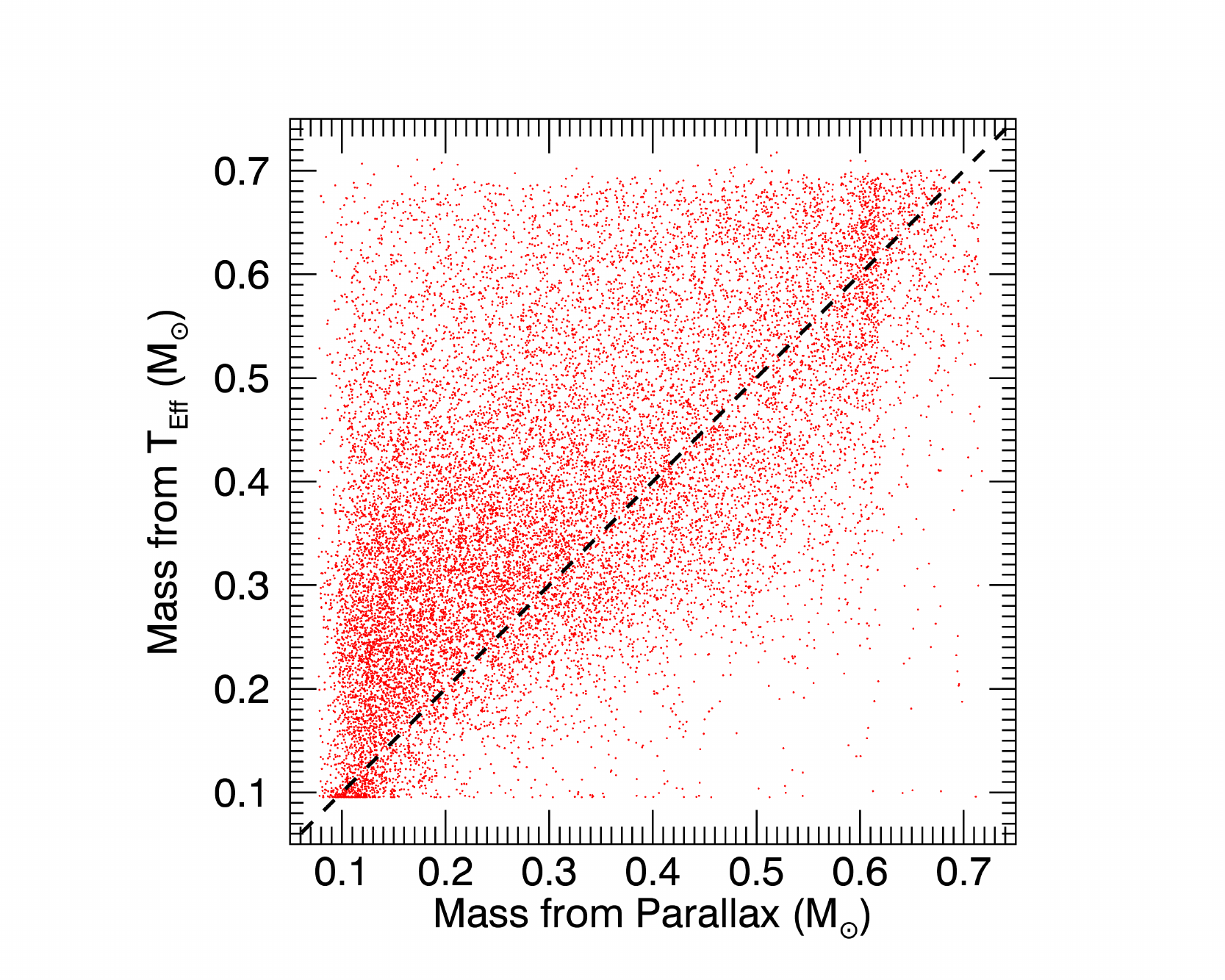}
    \includegraphics[width=\linewidth]{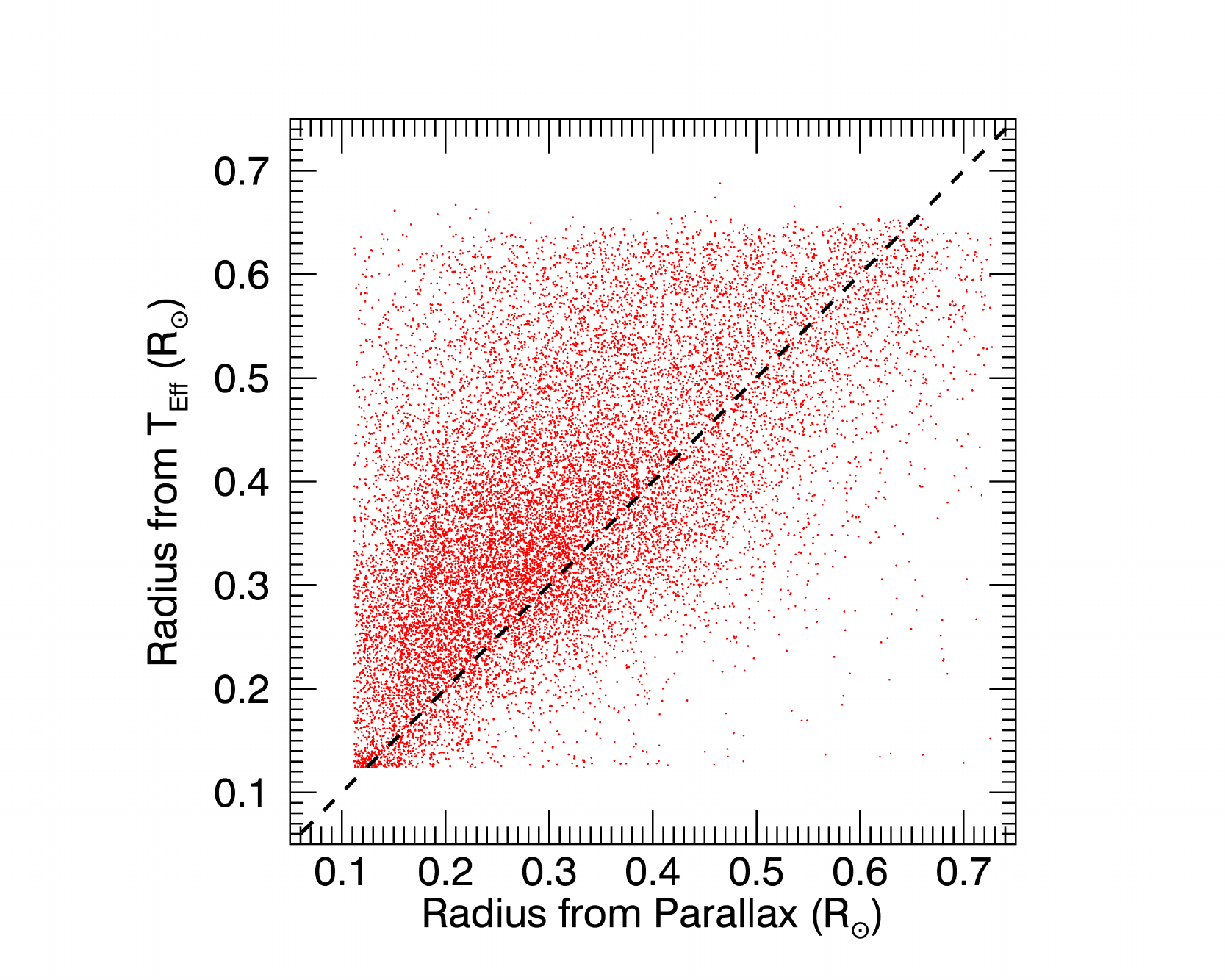}
    \caption{{\it Top}: Stellar mass determined from parallaxes vs. stellar mass determined from $T_{\rm Eff}$, for stars in the catalog with astrometric parallaxes from the literature.  {\it Bottom}: Same, but for stellar radius.\label{mass_radius}}
\end{figure}

To test the accuracy of the stellar parameters as determined from magnitudes alone, we use those cool stars with parallax observations.  Figure \ref{mass_radius} show the masses and radii as determined from trigonometric parallaxes, compared to the masses and radii as determined using magnitudes via the estimated effective temperature.  The root-mean-square deviation between the magnitude and parallax determinations is 0.145 $M_\Sun$ and 0.120 $R_\Sun$ for the mass and radius determinations, respectively.  The root-median-square deviation, a statistic less sensitive to outliers, between the magnitude and parallax determinations is 0.097 $M_\Sun$ and 0.081 $R_\Sun$ for the mass and radius determinations, respectively.  

It is clear from Figure \ref{mass_radius} that the mass and radius estimates from photometry alone have systematic errors, especially between 0.2 and 0.3 $M_\Sun$ and $R_\Sun$, where the photometric relations appear to over-predict the stellar masses and radii.  Such systematic errors are prevalent when using photometry alone to estimate stellar parameters of low-mass stars.  We note that the {\it Kepler} Input Catalog also faced systematic uncertainties in cool dwarf parameters, described \citep[][]{Brown2011} and measured via spectroscopy by \citet[][]{Muirhead2012a} and \citet[][]{Muirhead2014}.  Recently, \citet[][]{Dressing2017a}, \citet[][]{Dressing2017b}, \citet[][]{Martinez2017} and \citet[][]{Hirano2017} acquired spectra for cool dwarf planet hosts from NASA's K2 Mission, again revising stellar and planetary parameters originally determined by photometry.   We fully expect to revise the masses and radii of the objects in the Cool Dwarf Catalog as astrometric parallaxes become available from the Gaia Mission, as well as add additional objects primarily in the southern ecliptic hemisphere.

\subsection{Caveats and Warnings}
While the simplicity of the above methods for deriving properties of M dwarfs makes it easier to reproduce, and hence useful for the large catalog of stars required, there are a number of important caveats that one should consider when using these parameters;
\begin{itemize}
    \item All targets are assumed to be late K or M dwarfs. Reddened K stars, stars with inaccurate colors, or evolved stars with large or erroneous proper motions in the final sample will have faulty assigned stellar parameters. 
    \item Radii and masses derived from $M_{K_s}$ will be systematically large if the target is an unresolved binary. For near-equal mass binaries, this effect can result in derived radii as much as 30\% larger than the true value, significantly larger than the typical errors. Some of these systems could be identified by their position on a color-magnitude diagram (CMD). In practice, however, binaries are difficult to disentangle from metal-rich stars, due to the strong effect of metallicity on CMD position for M dwarfs \citep{Johnson2009,Schlaufman2010,Neves2011}, and employing such a correction might bias the sample in unintended ways. 
    \item Because metallicities are not known, masses and radii derived using \teff\ will be systematically too large for metal-poor stars and systematically too small those that are metal-rich. A small correction for this could be to include $J-H$ colors as a proxy for [Fe/H] in the \teff-$R_\star$ calculation, as was done when converting colors to \teff. However, a color-based metallicity correction to the radius-\teff\ relation has not been empirically calibrated to date. Therefore, we chose not to apply this correction.
    \item For parallax- or \teff-based methods, relations from \citet{Mann2015} are only valid over $-0.6<$[Fe/H]$<0.4$, and the \citet{Benedict2016} calibration sample is mostly near solar metallicity. For stars well below this range or with unusual abundance patterns (e.g., high C/O) the relations may be completely invalid. 
    \item There are significant systematic differences between different sources of photometry due to color-terms \citep[][]{Bessell2012, Mann2015b}. These terms are largest for the reddest stars, so this may be significant for M dwarfs in our sample.
    \item We assumed all stars are on the main sequence (MS). Extremely young stars ($\lesssim30$\,Myr) are likely a small fraction of our sample, and would be best handled using completely different techniques. Similarly, we assume stars are not being inflated by activity, whether through youth or external factors (e.g., a tight binary). The size of this effect is debated \citep[e.g.,][]{Kraus2011,Stassun2012,Mann2015,Mann2017b}, and we have no activity measurements across the majority of the sample, making it difficult to apply a correction. 
\end{itemize}

\section{Planet Yields}\label{sec:yield}
We estimated the anticipated planet yield from TESS observations of cool dwarfs by assigning planets to the stars using the occurrence rates derived from analyses of \Kepler data and determining the number of planets that could be detected by TESS. We began by constructing a fine grid in planet radius and orbital period space using the same boundaries as in Figure~11 of \citet{Dressing2015}. 


Next, we determined the transit depths and durations for planets at the centers of each grid cell in orbit around the stars in the cool dwarf sample.  We restricted our analysis to the 1,140,164 stars with assigned temperatures and masses in the range 0.08 $M_\Sun$ to 0.73 $M_\Sun$.  For each star/planet combination, we predicted the signal-to-noise ratio (SNR) for a single transit by comparing the expected transit depth to the noise predicted by the relations in S15. Our noise estimates incorporate shot noise, sky noise, readout noise, stellar noise, and a 60~ppm noise floor due to systematic effects, a conservative estimate.  Following S15, we included stellar noise by randomly assigning each star a noise level matching that of one of the 99 moderately bright ($12.5 < m_{\rm Kep} < 13.1$) Kepler cool dwarfs studied by \citet{Basri2013}. The distribution of assigned variabilities extends from 26 ppm to 1\% with a mean value of 326 ppm. Our noise estimates do not consider flux contamination due to nearby stars.

\begin{figure}[tbp]
    \centering
    \includegraphics[width=\linewidth]{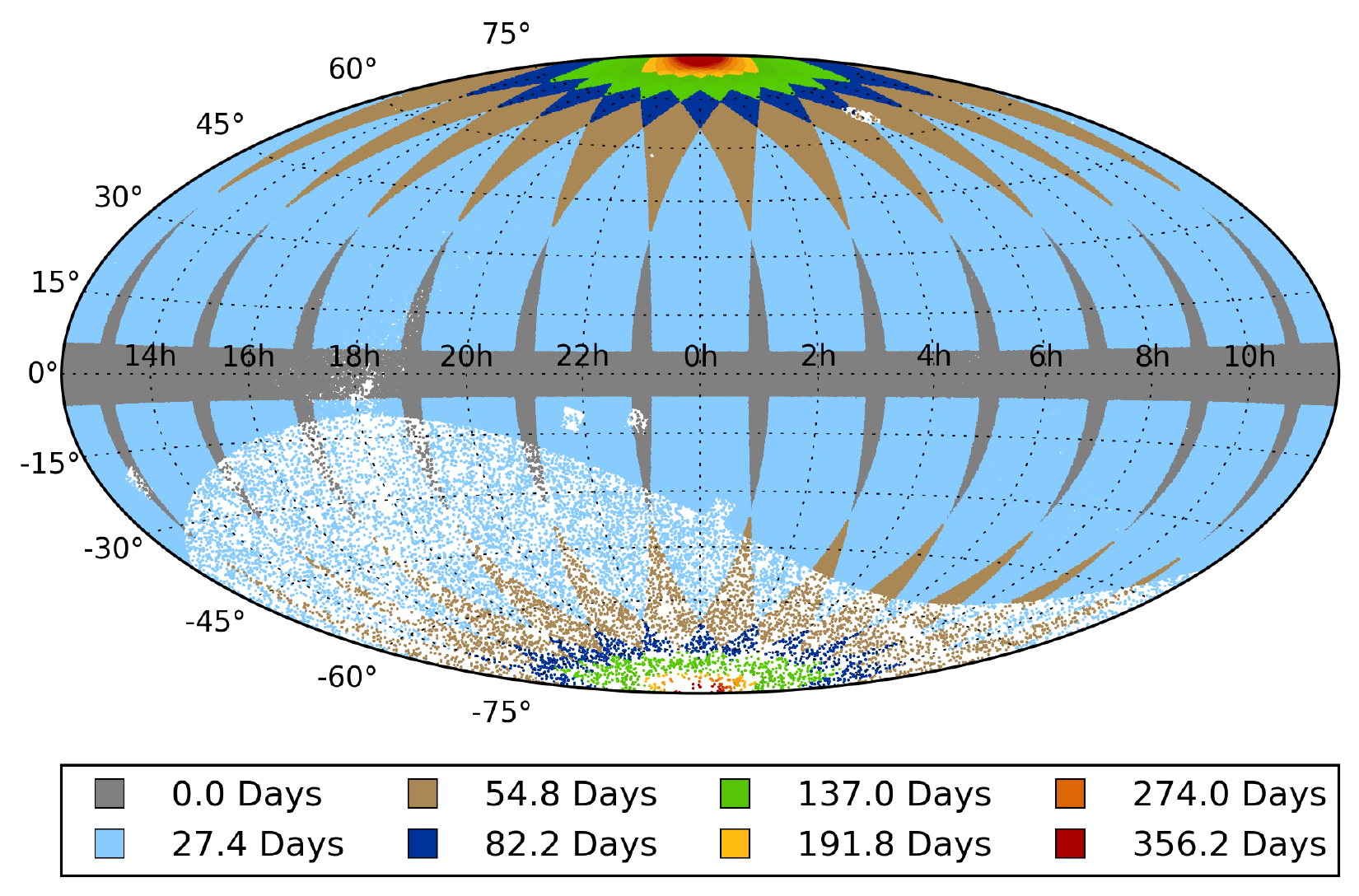}
    \caption{Map (in ecliptic coordinates) showing the number of days of data that could be obtained for each star in the cool dwarf catalog. Note that we assumed an arbitrary ecliptic longitude for the center of the first field.}
    \label{fig:survey_geometry}
\end{figure}

After estimating the single transit SNR (SNR$_{\rm single}$), we calculated the cumulative multiple transit SNR (SNR$_{\rm multi}$) by first determining the number of days that TESS could observe each star and then scaling SNR$_{\rm single}$ by the square root of the total number of transits. We estimated survey coverage by using a modified version of the tvguide tool developed by the TESS GI office \citep[][]{Mukai2017} assuming an arbitrary initial ecliptic longitude for the center of the first field (see Figure~\ref{fig:survey_geometry}).  We consider planets to be ``detected'' if  SNR$_{\rm multi}$ exceeds $7.1\sigma$, even if only one transit occurs within the observing window.

We account for the geometric consideration that not all planets will appear to transit. For each star, we generated a map of the occurrence rate of transiting planets by multiplying the occurrence rates from \citet{Dressing2015} by the geometric likelihood of transit $R_\star/a$ at the center of each grid cell. We then compute the number of detected planets per star by summing the occurrence rates of transiting planets in detectable grid cells (i.e., grid cells for which the SNR$_{\rm multi} > 7.1\sigma$). 


Finally, we estimated the total planet yield from the full cool dwarf population by adding the contributions from each star. In total, we anticipate that TESS would detect 2,136 planets with radii $0.5 R_\Earth< R_p <4R_\Earth$ and periods $0.5 < P < 200$~d if all stars in the cool dwarf catalog are monitored at 2-minute cadence when they are visible to TESS, assuming all objects are indeed M dwarfs. Subdividing the planets by radius, we anticipate that TESS would detect roughly  151 Earth-sized planets ($R_p <1.25R_\oplus$), 504 Super-Earths ($1.25R_\oplus < R_p <2R_\oplus$), and 1,481 sub-Neptunes ($2R_\oplus < R_p <4R_\oplus$) orbiting cool dwarfs. The typical TESS cool dwarf planet would be a $2.3 \, R_\oplus$ planet with a 7-day orbital period. However, we note that only 200,000 to 400,000 stars will be monitored in 2-minute cadence and that not all 2-minute targets will be cool dwarfs.  We discuss the selection of best objects for 2-minute cadence in Section \ref{twomin}.

\subsection{Prioritizing Targets for 2-minute cadence}\label{twomin}
The cool dwarf target list contains over 1 million stars, which is ten times larger than the set of stars that will be observed at 2-minute cadence. In order to investigate which cool dwarfs would benefit most from 2-minute cadence, we began by considering the dependence of the transit duration on host star properties. Following \citet{winn2010} and making the simplifying assumptions that (1) the planet is much smaller than the star; (2) the planet follows a circular orbit; (3) the orbital semimajor axis $a$ is much larger than the stellar radius $R_\star$; and (4) the planet transits directly across the center of the star, the approximate transit duration is

\begin{equation}
T \approx \frac{R_\star P}{\pi a}
\end{equation}
where $P$ is the planetary orbital period. Using Kepler's third law to rewrite $a$ in terms of $P$ and combining the stellar mass and radius terms into stellar density $\rho_\star$, the expression becomes
\begin{equation}
T \approx 13~{\rm h}\left(\frac{P}{\rm 1\, yr}\right)^{1/3}\left(\frac{\rho_\star}{\rho_\odot}\right)^{-1/3}
\label{eq:duration}
\end{equation}

If we then enforce the (arbitrary) requirement that TESS obtain at least $n$ data points during each transit event, then we find that 2-minute cadence observations are required for stars with densities 

\begin{equation}
\frac{\rho_{\star,threshold}}{\rho_\odot} \gtrsim  \left(\frac{1}{n} \frac{13~{\rm h}}{0.5~{\rm h}}\right)^3 \left(\frac{P}{\rm 1\, yr}\right) = \frac{48}{n^3}\left(\frac{P}{\rm 1\, d}\right)
\end{equation}

For a fiducial orbital period of 0.5~d, the critical stellar density required to obtain one point per transit is $\rho_{\star,threshold} = 24\rho_\odot = 34$~g~cm$^{-3}$, which roughly corresponds to stellar masses $\lesssim 0.15M_\odot$. If we instead change the fiducial orbital period to 2~days and require two points per transit, the critical stellar density decreases to 17~g~cm$^{-3}$, increasing the stellar mass cutoff to $\lesssim 0.27 M_\odot$. Clearly, the choice between 2-minute and 30-minute cadence is strongly dependent on the targeted orbital period range. Maximizing the yield of ultra-short period planets would require that the majority of cool dwarfs be observed at 2-minute cadence.

Given the practical limitations on the number of cool dwarfs that can actually be monitored at 2-minute cadence, one strategy for selecting 2-minute cadence targets would be to prioritize cool dwarfs that are bright enough for follow-up observations. While there are 326,039 cool dwarfs with densities higher than 17~g~cm$^{-3}$, only 5,243 have $T < 13$. A fraction of these stars are likely to be unobservable due to gaps between detectors and proximity to extremely bright stars, further reducing the required number of 2-minute cadence pixels.

\subsection{Considering Characterization}
Even if 2-minute cadence is not required to detect planets orbiting larger cool dwarfs or planets with longer orbital periods, acquiring observations at shorter cadence significantly improves the precision of stellar parameter estimates. For instance, the impact parameter can be constrained much more precisely when the shape of the transit profile is captured. In turn, narrowing the allowed range of impact parameter improves the constraints on planet radius. 

Transit durations are also much easier to measure with high cadence photometry. Accordingly, estimating the orbital periods of single-transit events will be less challenging if those transits happen to be observed at 2-minute cadence. Due to their longer orbital periods,  singly transiting planets are likely to be some of the most enticing targets from a planetary habitability perspective, so having the ability to determine their ephemerides and recover their transit windows would be advantageous. Such systems are significantly more compelling if they orbit stars bright enough for atmospheric studies or planetary mass measurement, presenting another justification for prioritizing the cool dwarf target list by host star magnitude and follow-up potential as well as planet detectability.

The final choices of how many cool dwarfs should be observed at 2-minute cadence and how those targets should be selected are beyond the scope of this paper and will be decided by the TESS mission and Guest Investigator Office. In order to help inform that decision, we now study how four extreme choices of prioritization schemes might influence the yield of small planets orbiting cool dwarfs. For each simulation, we select the best 25,000 stars according to the following criteria:   
\begin{enumerate}
    \item \textbf{Dense Stars}: As shown in Equation~\ref{eq:duration}, planets orbiting denser stars have shorter transit durations. This prioritization scheme aims to maximize the yield of (ultra) short period planets by preferentially reserving 2-minute cadence observations for the densest cool dwarfs.
    \item \textbf{Bright Dense Stars}: Selecting targets by density alone strongly biases the sample toward faint late M dwarfs that are challenging targets for planet detection and follow-up observations due to low photon counts. This scheme uses the same density ranking as the ``Dense Stars'' scheme but requires that all stars have $T < 13$.
    \item \textbf{Bright Stars}: If the goal of detecting planets with TESS is to identify a sample of planets amenable to follow-up mass measurement and atmospheric characterization, then bright host stars are advantageous. This scheme ranks targets by TESS magnitude. 
    \item \textbf{``Easy'' Stars}: Selecting easier search targets is attractive because search incompleteness is lower and a higher planet yield could be detected using a smaller target list.  This prioritization scheme orders targets by the cumulative signal-to-noise ratio expected due to multiple transits of an Earth-radius planet with a period of 7 days. 
\end{enumerate}

\begin{figure}[tbp]
    \centering
     \includegraphics[width=1\linewidth]{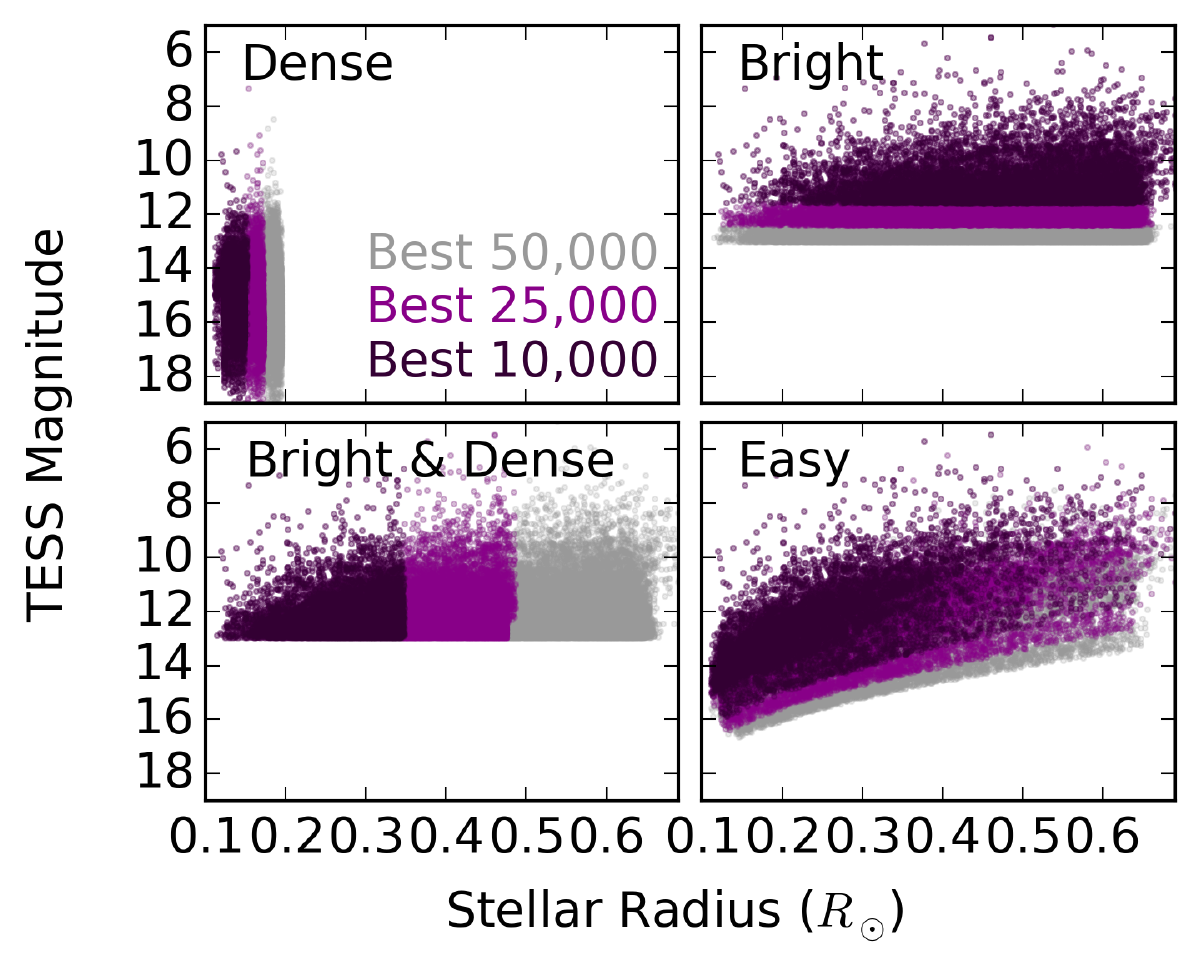}\\
    \caption{Comparison of the brightnesses and radii of stars selected for observation at 2-minute cadence using various prioritization schemes: ``Dense'' (top left), ``Bright'' (top right), ``Bright \& Dense'' (bottom left) and ``Easy'' (bottom right). Within each panel, the best 10,000 stars are shown in dark purple, stars within the best 25,000 are shown in purple, and stars within the best 50,000 are shown in gray.} \label{fig:stars_selected}
\end{figure}

Overall, the TESS mission is expected to monitor 200,000 to 400,000 stars at 2-minute cadence, so selecting 25,000 stars is akin to devoting 6 to 13\% of the observing time to cool dwarfs. In addition to these baseline simulations, we also run ``bottom-heavy'' and ``top-heavy'' simulations in which we select 50,000 and 10,000 cool dwarfs per mission. We display the stars selected for each simulation in Figure~\ref{fig:stars_selected}. As expected, the ``Dense'' and ``Bright'' samples contain small stars and bright stars, respectively. The ``Dense \& Bright'' sample spans approximately the same magnitude range as the ``Bright'' sample, but is biased toward smaller stars. The ``Easy'' sample contains a mixture of smaller, fainter stars and larger, brighter stars.

\begin{figure}[tbp]
    \centering
     \includegraphics[width=1\linewidth]{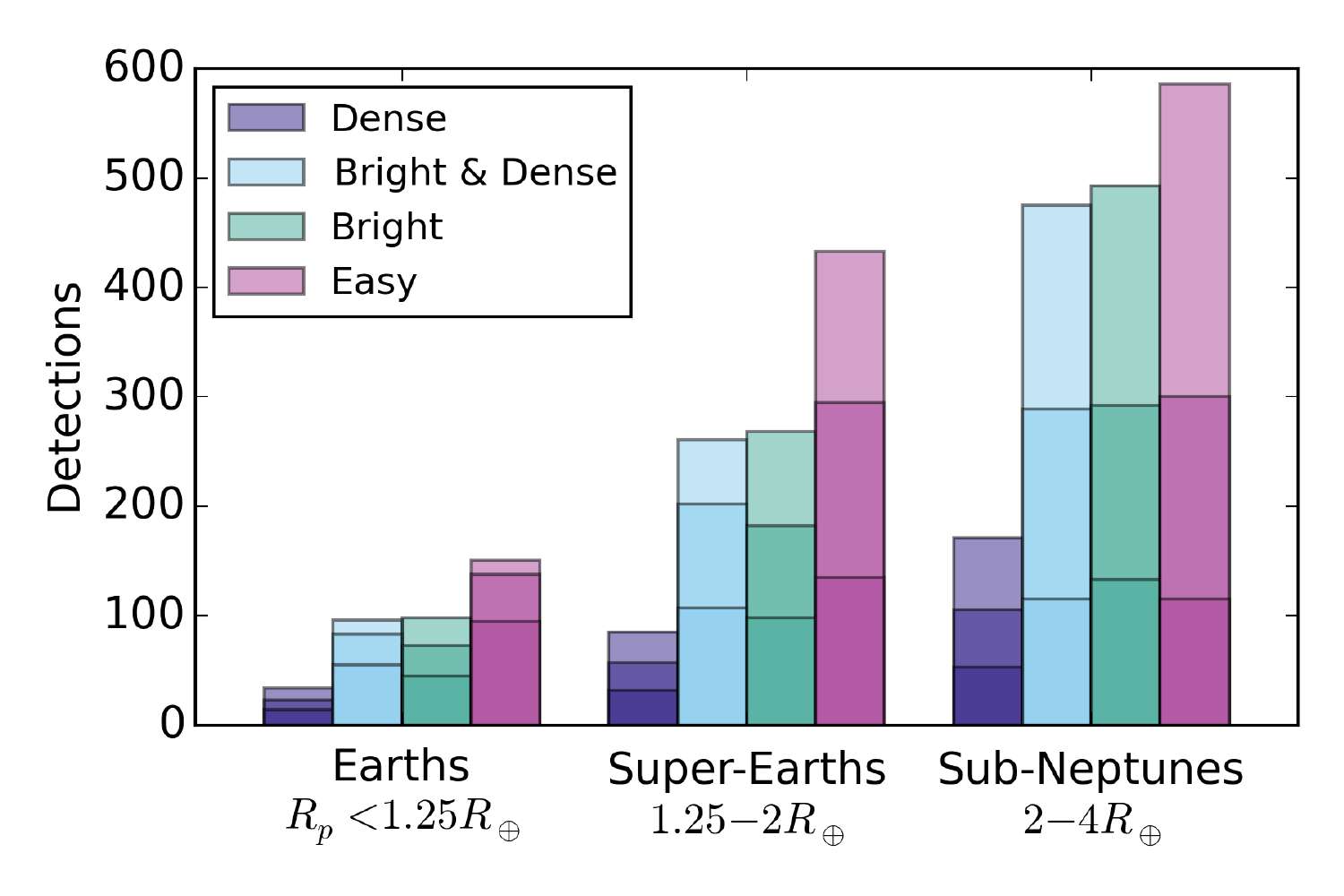}\\
    \caption{Number of expected detections as a function of planet radius for four notional target prioritization schemes. For each scheme, the short, medium, and tall bars indicate the number of planets that would be detected if the best 10,000, 25,000, or 50,000 stars were selected according to the chosen prioritization scheme.} \label{fig:yield_comparison}
\end{figure}

Comparing the resulting planet yields, we found that the choice of prioritization scheme has a dramatic effect on the number of detected planets. We display the resulting size distributions of detected planets for all four prioritization schemes in Figure~\ref{fig:yield_comparison}. In general, the ``Dense Star'' scheme recovers far fewer planets than all other proposed schemes: observing the 10,000 densest stars would yield detections of only 14 Earth-sized planets ($R_p < 1.25 R_\oplus$), 32 Super-Earths ($1.25-2 R_\oplus$, and 53~Sub-Neptunes ($2-4 R_\oplus$). For comparison, observing the 10,000 brightest stars or the 10,000 ``easiest'' stars would yield roughly 45 Earths, 98 Super-Earths, and 133 Sub-Neptunes or 95 Earths, 135 Super-Earths, and 115 Sub-Neptunes, respectively. Considering that the ``Easy'' scheme selects targets based on small planet detectability, the result that the ``Easy'' scheme finds the highest number of Earths is not surprising. The ``Bright'' survey performs better than the ``Easy'' survey in terms of the number of Sub-Neptune detections (133 planets versus 115 if 10,000 stars are observed), but the ``Easy'' survey detects a larger population of smaller planets. 

The relatively poor yield from the ``Dense Star'' scheme is likely due to the faintness of the densest stars. Although short period planets orbiting those stars would transit very quickly, there is little benefit to observing those stars at 2-minute cadence if the photon counts are too low to permit planet detection. We therefore recommend against using stellar density alone to select 2-minute cadence targets. However, stellar density may be a useful selection criterion when used in combination with other metrics. For instance, we find that a survey of the 10,000 best ``Bright \& Dense'' stars would detect 55~Earths, 107~Super-Earths, and 115~Sub-Neptunes.

\begin{figure}[tbp]
    \centering
     \includegraphics[width=1\linewidth]{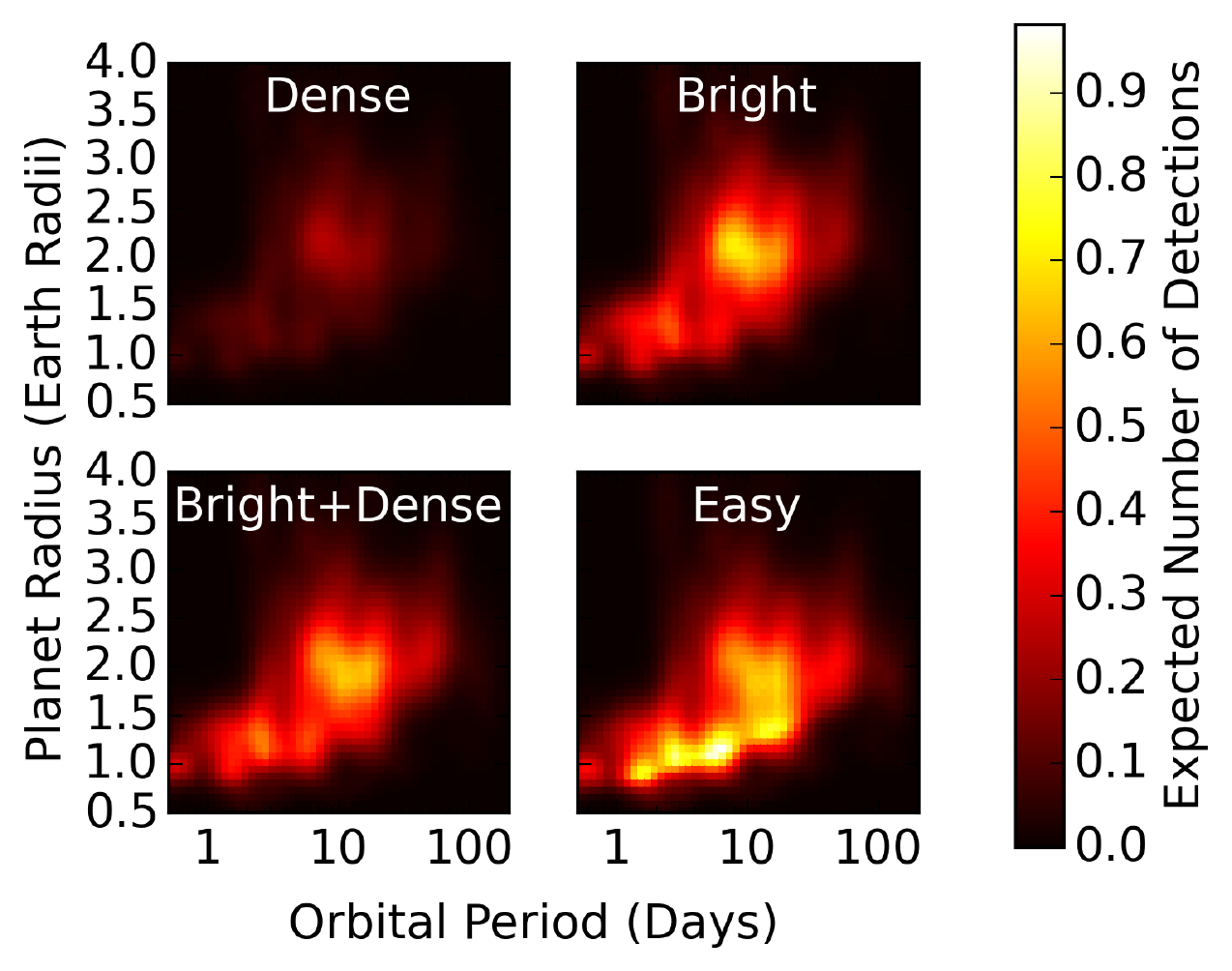}\\
    \caption{Number of expected detections as a function of planet radius and orbital period for four notional target prioritization schemes. For each scheme, the color indicates the expected number of detections per grid cell. We selected 10,000 target stars per simulation.} \label{fig:small_yields}
\end{figure}

The differences in the populations of detected planets are highlighted in Figure~\ref{fig:small_yields}, which shows heatmaps of the detected planet yield as a function of planet radius and orbital period. The trace for the ``Dense'' survey is barely noticeable, demonstrating that the stars with the highest densities are challenging transit targets. The heatmap for the ``Bright'' survey displays a clear pileup of planets with periods near 10~days and sizes of roughly $2 R_\oplus$. These planets would be viable targets for atmospheric characterization and possibly mass measurement. The same planet pileup is noticeable in the ``Bright \& Dense'' heatmap, but the overall distribution of planets is flatter. Unlike the other distributions, the heatmap for the ``Easy'' survey displays an enticing ridge of planet detections at small radii ($0.8-1.5 R_\oplus$) and short orbital periods (1.5 - 20~d). Detecting planets like these would be an excellent opportunity to determine the prevalence and composition of smaller planets.

\section{Discussion}\label{sec:discussion}

In this article we presented an all sky catalog of cool dwarf targets for the TESS Input Catalog based largely on archival photometry, parallaxes (where available) and reduced proper motions.  We estimated the stellar properties of the cool dwarfs in the catalog based on archival relationships between color, temperature, stellar mass and stellar radius.  We also estimated each star's TESS magnitude ($T$), for the purpose of estimating the ability to detect transiting planets around each star.

We purposefully ignored the role of binarity and interstellar reddening on the properties listed in the catalog.  We ignore binarity because it is difficult to determine the binarity of the stars in the sample with the archival data and it is unclear how this should affect the choice of exoplanet search targets.  We ignored reddening as it was calculated to have a marginal affect on the reported stellar properties. 

Lastly, we considered several prioritization schemes to determine which of the stars in the cool dwarf catalog would benefit significantly from two-minute observations.  We used results from NASA's Kepler Mission to estimate the planet population around the stars in the cool dwarf catalog.  We find that prioritizing the targets based on stellar density would result in far fewer planet detections than prioritizing based on signal-to-noise or star brightness.  We find that prioritizing stars based on signal-to-noise (``Easy'') or star brightness (``Bright'') results in a similar number of detections, but that the Easy yield would contain more small planets.  Our yield simulations did not consider the effects of flux contamination due to nearby stars or the coplanarity of multi-planet systems. See \citet{Ballard2018} for a detailed analysis of how the low mutual inclinations of planets in multi-planet systems could increase the planet yield.

With the anticipated release of parallax observations from the Gaia Mission, many of the stellar properties in this catalog will be revised.  Until then, however, this catalog serves as a catalog of cool dwarf targets for the TESS Mission, including the primary science mission and Guest Observer programs.  The cool dwarf catalog has been incorporated into the TESS Input Catalog and is available for download online.

\acknowledgements

We would like to thank George Ricker, Keivan Stassun, Joshua Pepper, and the entire TESS Star Properties Working Group for encouraging this work in support of the TESS Mission.  We would like to thank Michael Cushing for pointing out the importance of coefficient covariances in mass-luminosity relations.  We would like to thank Josh Winn and Luke Bouma for sharing their insight on planet yields.

P.S.M acknowledges support from the NASA Exoplanet Research Program (XRP) under Grant No. NNX15AG08G issued through the Science Mission Directorate.  B.R-A. acknowledges the support from CONICYT PAI/Concurso Nacional Inserci{\'o}n en la Academia, Convocatoria 2015 79150050.  This work was performed in part under contract with the Jet Propulsion Laboratory (JPL) funded by NASA through the Sagan Fellowship Program executed by the NASA Exoplanet Science Institute.  This research has made use of the NASA Exoplanet Archive, which is operated by the California Institute of Technology, under contract with the National Aeronautics and Space Administration under the Exoplanet Exploration Program \citep[][]{Akeson2013}.

This work has made use of data from the European Space Agency (ESA)
mission {\it Gaia} (\url{https://www.cosmos.esa.int/gaia}), processed by
the {\it Gaia} Data Processing and Analysis Consortium (DPAC,
\url{https://www.cosmos.esa.int/web/gaia/dpac/consortium}). Funding
for the DPAC has been provided by national institutions, in particular
the institutions participating in the {\it Gaia} Multilateral Agreement.

\vspace{5mm}

\software{HEALPix \citep{Gorski2005}, 
    TRILEGAL \citep{Girardi2005},
    VESPA \citep{Morton2012,Morton2015}
}


\bibliographystyle{aasjournal.bst}
\bibliography{bibfile}

\end{document}